# Understanding Hot-Electron Generation and Plasmon Relaxation in Metal Nanocrystals: Quantum and Classical Mechanisms

Lucas V. Besteiro,[1*] Xiang-Tian Kong,[1,2] Zhiming Wang,[2] Gregory Hartland,[3] Alexander O. Govorov[1*]

[1] *Department of Physics and Astronomy, Ohio University, Athens OH 45701*

[2] *Institute of Fundamental and Frontier Sciences and State Key Laboratory of Electronic Thin Films and Integrated Devices, University of Electronic Science and Technology of China, Chengdu 610054, China*

[3] *Department of Chemistry and Biochemistry, University of Notre Dame, Notre Dame, IN 46556-5670*

[*] E-mails: lvbesteiro@gmail.com; govorov@ohio.edu

**Abstract**: Generation of energetic (hot) electrons is an intrinsic property of any plasmonic nanostructure under illumination. Simultaneously, a striking advantage of metal nanocrystals over semiconductors lies in their very large absorption cross sections. Therefore, metal nanostructures with strong and tailored plasmonic resonances are very attractive for photocatalytic applications in which excited electrons play an important role. However, the central questions in the problem of plasmonic hot electrons are the number of optically-excited energetic electrons in a nanocrystal and how to extract such electrons. Here we develop a theory describing the generation rates and the energy-distributions of hot electrons in nanocrystals with various geometries. In our theory, hot electrons are generated due to surfaces and hot spots. As expected, the formalism predicts that large optically-excited nanocrystals show the excitation of mostly low-energy Drude electrons, whereas plasmons in small nanocrystals involve mostly high-energy (hot) electrons. We obtain analytical expressions for the distribution functions of excited carriers for simple shapes. For complex shapes with hot spots and for small quantum nanocrystals, our results are computational. By looking at the energy distributions of electrons in an optically-excited nanocrystal, we see how the quantum many-body state in small particles evolves towards the classical state described by the Drude model when increasing nanocrystal size. We show that the rate of surface decay of plasmons in nanocrystals is directly related to the rate of generation of hot electrons. Based on a detailed many-body theory involving kinetic coefficients, we formulate a simple scheme describing how the plasmon in a nanocrystal dephases over time. In most nanocrystals, the main decay mechanisms of a plasmon are the Drude friction-like process and the interband electron-hole excitation, and the secondary path comes from generation of hot electrons due to surfaces and electromagnetic hot spots. The hot-electron path strongly depends on the material system and on its shape. Correspondingly, the efficiency of hot-electron production in a nanocrystal strongly varies with size, shape and material. The results in the paper can be used to guide the design of plasmonic nanomaterials for photochemistry and photodetectors.

# Introduction.

The generation of energetic (hot) electrons in plasmonic nanostructures is currently attracting lots of attention.[1–27] The motivations for this research topic are both fundamental and applied. It is interesting to learn more about the complexity of the plasmon's wavefunction in a nanocrystal (**NC**). Such wave functions involve many interacting electrons. One convenient way to see the structure of the many-body state of a plasmon is to compute the energy distribution of electrons.[28,29] Experimentally, this topic mainly concerns two types of applications: nanomaterials for enhanced photochemistry[1,2,5–20,30] and photodetectors. [3,4,21–27]

Current literature has a large number of theoretical publications on hot plasmonic electrons. A variety of quantum methods has been applied to the problem, such as Fermi's golden rule, an approach of non-equilibrium Green functions, quantum kinetic equations, etc.[28,29,31–39] The distribution of hot carriers has been calculated in the time-dependent pulse-excitation regime[39] or in the steady-state regime under continuous wave (CW) illumination.[28,29] Furthermore, theoretical models were developed for hot-electron photocurrents across metal-semiconductor barriers.[32,34,35] Our choice for the treatment of hot electrons involves a perturbative approach to solve the master equation for the electronic density matrix with a few relaxation times.[28,29,40–42] We employ the kinetic quantum approach of the master equation for the following reasons: (1) it provides a realistic description of electronic and plasmonic kinetics in real metals; (2) it affords us the ability to directly calculate the non-equilibrium energy distributions of hot plasmonic electrons generated in typical metal nanocrystals; (3) this method involves two or three relaxation times and allows us to reproduce dynamics of the 3D systems in the limit of large sizes. In contrast to the quantum field theory descriptions of plasmonic excitations,[43–45] our approach gives access to the non-equilibrium populations of single-electron quantum states in a plasmonic wave.

It was noticed in theoretical papers[40,42,46] that plasmonic NCs with hot spots and complex shapes may generate unusually large numbers of hot electrons. Plasmonic hot spots are small regions of space in metal NCs with complex shapes where the incident electromagnetic field becomes strongly enhanced due to the geometry of the system. Examples of NCs with hot spots include nanocubes, dimers, nanostars, etc.[47] One can expect enhanced photochemical processes in hot spot regions of a plasmonic nanostructure and indeed such hot-spot induced processes were recorded experimentally.[48,49] Later the hot-spot effect was invoked to explain the photochemical performance of NCs of different shapes[50,51] and also a plasmonic experiment with an unusual ultra-fast response.[52] According to our previous theoretical modeling,[40,42,46] two mechanisms can contribute to the enhanced generation of energetic electrons in NCs with hot spots: (1) The effect of field enhancement, which is typical for the plasmonic systems, and (2) the effect of non-conservation of linear momentum of electrons owing to surfaces and hot spot regions. The first mechanism is a classical phenomenon, whereas the second one needs a quantum treatment.

Here we will describe the quantum properties of plasmons in confined geometries and focus on the process of generation of energetic (hot) electrons. Figure 1a shows the plasmonic geometries and their inner field enhancement. Our goals in this study are to find a simple and transparent description for the effect of hot-electron generation and to relate this effect with the decay rate of the plasmon.

The present study has a number of new, important developments as compared to our previous papers on this topic.[28,29,40–42,46] (1) In this study, we introduce the quantum detailed-balance equation for the density matrix and then treat the hot-electron processes using the spectral rate of generation. This allows us to obtain results that are independent of relaxation mechanisms. Such spectral rates can be used to understand and guide many current experiments in the field. We note that our previous papers were focused on the calculation of steady-state populations of hot electrons that strongly depend on relaxation mechanisms. (2) For spherical NCs, we obtain simple analytical results for the hot-electron rates and efficiencies that can be used by many experimental groups working in the field. (3) We show how to treat computationally hot electrons in NCs with arbitrary shapes; such semi-analytical equations provide a way to compute intraband plasmonic excitations in experimental nanostructures with complex geometry, which has been a real challenge so far. For example, the theory developed here is fully applicable for a description of the plasmonic hot-electron phenomena in infrared metastructures and super-absorbers.[26,52] (4) The paper describes the connection between the hot-electron generation rates and the enhanced plasmon decay. (5) This study naturally generalizes Kreibig's theory of surface scattering to the case of NCs with hot spots. (6) The last part of this paper concerns the effect of enhanced hot-electron generation in plasmonic NCs with complex shapes hot spots. In particular, this paper describes how to treat two types of hot spots and how to predict the number of excited carriers in such complex plasmonic systems.

Our study focusses on the intraband transitions and treats electrons as free particles. The intraband transitions in typical spherical NCs made of gold or silver produce hot electrons with high energies which are able to pass through a Schottky barrier and trigger surface chemistry. In contrast, interband transitions for the typical spherical NCs produce electrons with relatively small energies. The spectra of interband hot carriers should be calculated using the band structure of a bulk crystal; such calculations can be found, for example, in ref 41. Importantly, the interband transitions are not active in the long wavelength intervals. For example, the interband transitions in the gold NCs are not active at wavelengths > 600 nm. Therefore, our formalism for the intraband transitions gives the complete picture of the hot-electron generation for a variety of experimentally-relevant nanostructures with plasmonic resonances in the red and infrared spectral intervals, such as nanorods, nanostars,[46,50] platelets and planar metamaterials.[26,52]

This paper adopts the following structure. We will first provide an overview of our theoretical perspective. Then we will describe in detail the quantum formalism to compute hot-electron generation. The following part will include the non-equilibrium distribution of carriers, which depends on the choice of the relaxation operator. A later section will provide details of the numerical calculations for simple NCs without hot spots and discuss the relevance of material systems and relaxation mechanisms. We finally extend the discussion to NCs with hot spots, including complex shapes such as nanocubes and nanosphere dimers, before concluding the paper.

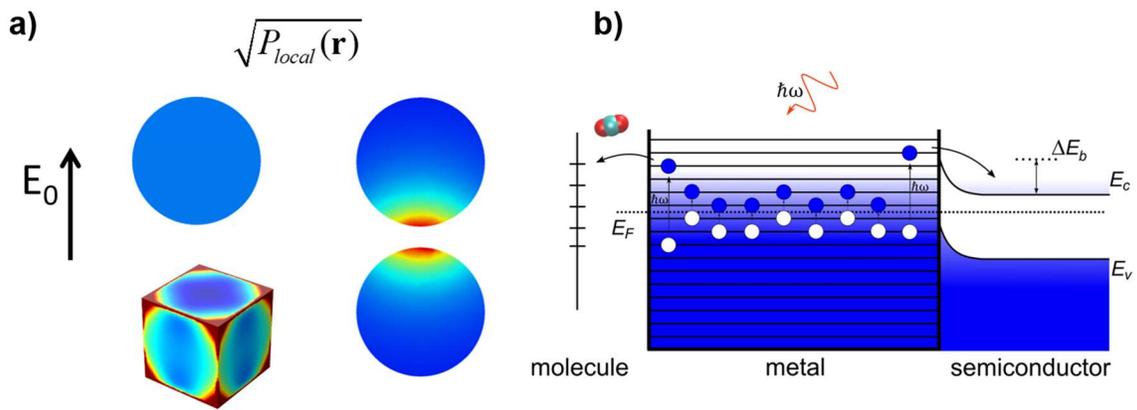

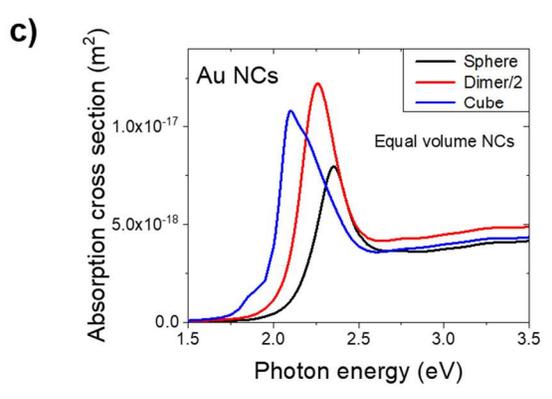
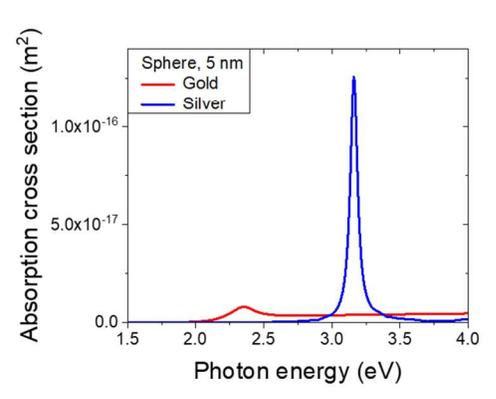

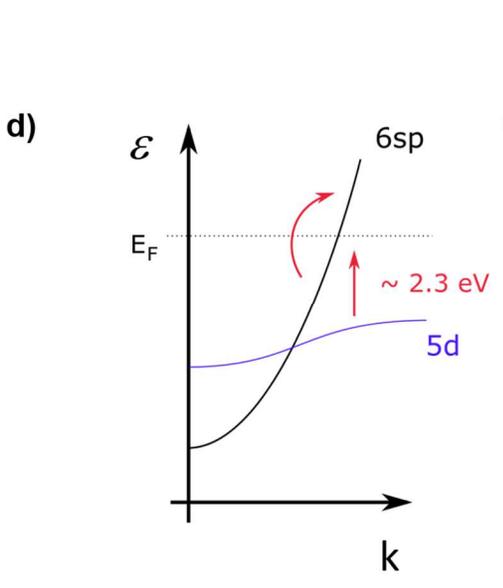
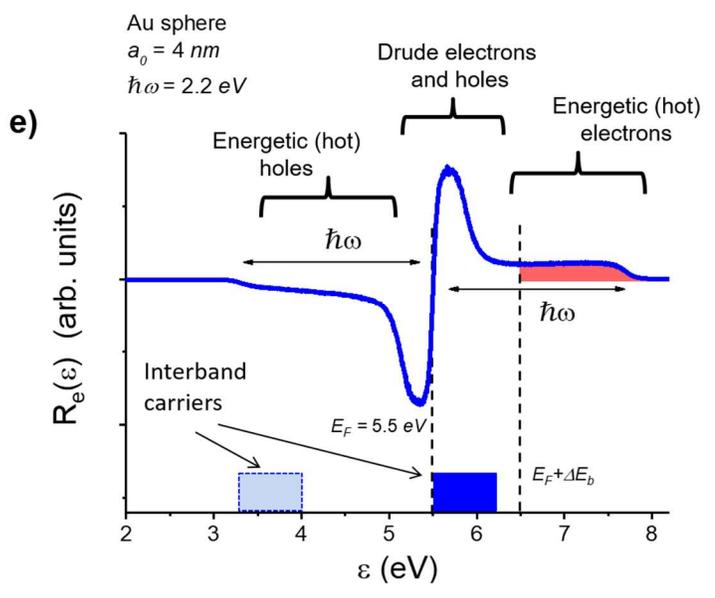

**Figure 1. (a)** Color maps of the electric-field enhancement inside nanocrystals. **(b)** Diagram showing single-electron excitations in a localized plasmon wave and the ways to extract hot electrons. **(c)** Absorption cross sections for Au nanocrystals with three different geometrical configurations and the comparison of absorption for Au and Ag nanospheres; volumes of all NCs are taken equal with an effective radius $R_{eff} = 2.5 nm$. **(d)** Simplified band diagram of gold and an illustration for interband and interband transitions. **(e)** Typical spectrum for the rate of electron generation in a localized plasmon wave in a metal NC. Two types of excited intraband carriers can be observed, regarded as Drude carriers and hot carriers. The two blue regions correspond to the intervals of generation of interband holes and electrons in gold.

## General considerations on hot electron generation and plasmonic dephasing.

Figure 1 illustrates classical and quantum properties of plasmonic NCs. In particular, Figure 1c shows that plasmonic spectra of metal NCs are strongly material- and shape-dependent. In the typical plasmonic metals (gold and silver), the electronic band structure includes two types of bands, the sp-band and the d-bands (Figure 1d). Therefore, illumination generates two types of excited carriers, promoted through intraband and interband transitions. In this paper, we will mostly deal with the intraband transitions, since they are able to generate excited electrons with high energy above the Fermi level. In Figure 1e, we show the spectral rate of intraband generation that appears in the interval of electronic energies $E_F + \hbar\omega > \varepsilon > E_F - \hbar\omega$. Electrons created by the interband transitions cannot typically reach energies as high as $E_F + \hbar\omega$. In Figure 1e, we show the interband generation as a red area. More information about the generation of interband electrons and holes can be found in ref 41 and also in the Supporting Information.

Figure 1e shows the typical spectrum for the rate of intraband generation of excited electrons over energy in a plasmonic NC in the steady-state (CW) illumination regime. We see that, above the Fermi level ($\varepsilon > E_F$, with $E_F = 5.5\ eV$ for gold), the rate of generation $R_e(\varepsilon)$ is positive, whereas, below the Fermi level, it is negative. Thus, we regard the excited carriers above the Fermi level as plasmonic electrons and the empty states below the Fermi level as holes. In the following sections we will describe excited carriers using the following spectral functions

$$R_e(\varepsilon) = \frac{d^2 N_e}{dt d\varepsilon},$$
$$\delta n(\varepsilon) = \frac{dN_e}{d\varepsilon}, \quad (1)$$

where $N_e$ is the number of excited electrons (i.e. electrons with energies above the Fermi level) in a plasmonic wave in a NC. The first function, $R_e(\varepsilon)$, provides us with the spectral information on the rate of generation of carriers. The second function $\delta n(\varepsilon)$ describes the spectral distribution of the carrier population in the steady state under CW illumination.

An optically-driven plasmonic NC has two types of excited intraband carriers (Figure 1d): Low-energy carriers forming coherent plasmonic currents and high-energy (hot) electrons excited primarily through surface effects and hot spots. The first type of carriers and the related electric currents are well described by the classical Drude model. These carriers have low energies and create frictional (Joule) heating in a NC. The second type of carriers comprises energetic (hot) electrons and holes, having an excitation energy up to $\hbar\omega$, where $\omega$ is the photon frequency. The generation of such hot carriers is a quantum effect and appears due to the non-conservation of linear momentum of electrons near surfaces and in hot spots. Then, the total rate of generation of carriers in a NC is given by the integral over energy:

$$Rate_{tot} = \int_{E_F}^{\infty} R_e(\varepsilon) \cdot d\varepsilon = \frac{dN_e}{dt}. \qquad (2)$$

Since we have two types of carriers, the total rate of generation of electron generation can be naturally split into two parts:

$$Rate_{tot} = Rate_{low-energy} + Rate_{high-energy}.$$

In a later section, we show that

$$\frac{Rate_{high-energy}}{Rate_{low-energy,Drude}} = const \cdot \frac{l_{mfp}}{a_0} \frac{v_F/a_0}{\omega}, \qquad (3)$$

where $\omega$ is the optical frequency, $l_{mfp}$ is the mean free path of an electron, $a_0$ and $v_F$ are the NC size and the Fermi velocity of metal, respectively.

Now we can look at the plasmonic relaxation mechanisms in a NC, where the plasmonic dynamics are intrinsically related to the generation of low-energy and high-energy carriers. In Figure 2, we illustrate schematically the plasmonic relaxation processes. It is well established that the rate of decay of a plasmon in a simple NC has the following contributions:[53,54]

$$\gamma_{plasmons} = \gamma_b + \gamma_{surface}, \qquad (4)$$

where $\gamma_b$ is the bulk contribution that includes the scattering of electrons by phonons and defects, and the term $\gamma_{surface}$ comes from electronic collisions with the surface in a NC, and thus depends on the NC size and geometry. The bulk term can be further split into two parts:

$$\gamma_b = \gamma_{Drude} + \gamma_{inter\text{-}band}, \qquad (5)$$

where $\gamma_{Drude}$ is the broadening coming from the Drude part of the dielectric function (see Supporting Information for details). In gold and silver, this term comes from the scattering of electrons by phonons and defects within the sp-band. The rate $\gamma_{inter\text{-}band}$ comes from the inter-band electron-hole transitions that can be excited by the localized plasmon or from direct photon absorption. The two terms in eq 5 naturally appear from the expression of the bulk dielectric constant of a metal, which can be written as a sum of the Drude and inter-band terms.[55] It is important to note that we do not consider in eq 4 the contributions from radiative processes, which are only important for relatively large NCs.[53]

The most interesting contribution in eq 4 is the so-called Kreibig's term [54,56–58] or surface scattering term:

$$\gamma_{surface} = A \cdot \frac{v_F}{a_0}, \qquad (6)$$

where the numerical constant $A$ depends on the shape of the NC. The term $\gamma_{surface}$ comes from the decay of plasmons into electron-hole pairs near the surface of a NC. One recent review focusing on this plasmonic decay mechanism can be found in ref [54]. In refs [28,29,40,42], the surface scattering process was identified as an efficient mechanism of generation of hot electrons in NCs with small sizes and hot spots. Such generation of hot carriers occurs in NCs because the linear momentum of electron is not conserved near the surface of a NC and, therefore, an electron can pick up the photon quantum, with energy $\hbar\omega$. This mechanism of generation of hot electrons in NCs is also strongly shape-dependent.[40,42,46]

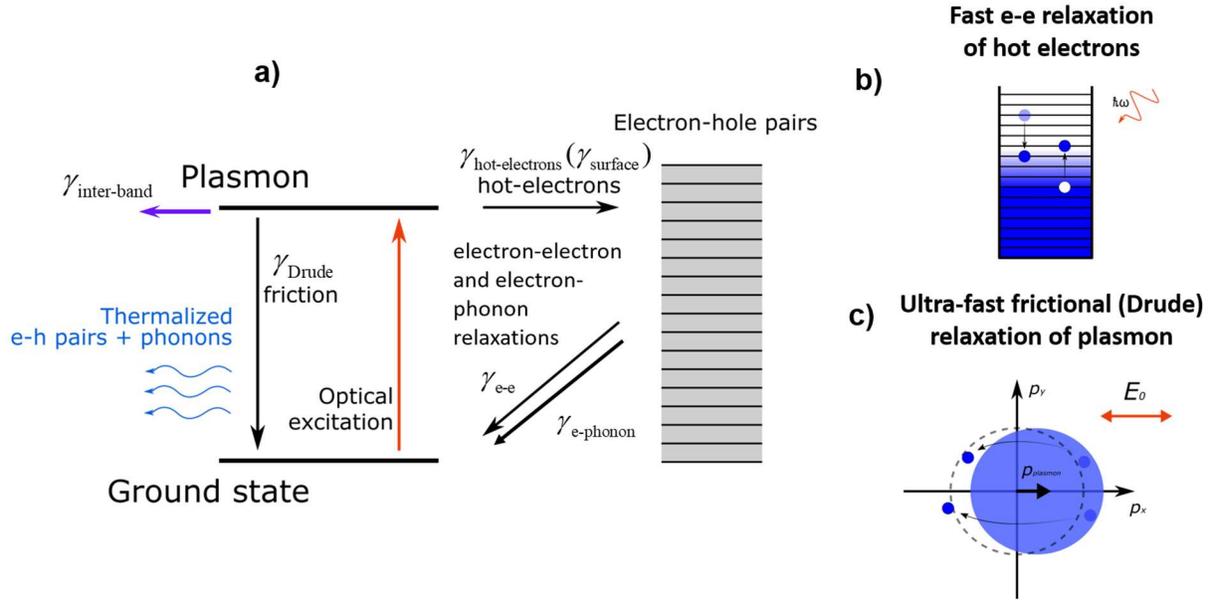

**Figure 2. (a)** Diagram of the processes of plasmon decay. The text explains in detail the relaxation channels shown in this panel. **(b,c)** Illustrations for two important mechanisms of relaxation: fast energy relaxation of hot electrons due to e-e scattering (b) and ultra-fast momentum relaxation of plasmon due to the frictional Drude force.

As one can see from Figure 2a, a plasmon decays into high- and low-energy electrons and holes in the Fermi sea. In the following step, these hot and warm electrons emit phonons and locally increase the lattice temperature.[59,60]

Having described the major parameters that characterize the plasmon decay and dephasing, we can now connect these processes with the generation of hot carriers. There is a simple relation between the generation rate $Rate_{high-energy}$, as defined above (eq 3), and the shape-dependent plasmonic relaxation parameter,

$$\gamma_{surface} = \frac{\hbar\omega \cdot Rate_{high-energy}}{E_{plasmon}} \qquad (7)$$

where the rate of high-energy electrons, $Rate_{high-energy}$, was defined above, and $E_{plasmon}$ is the energy stored in the coherent motion of the electrons in a localized plasmonic wave. This energy is composed of the kinetic and potential contributions:

$$E_{plasmon} = \int n_0(r) \frac{m_e v^2(r)}{2} dV + \frac{1}{2}\int \rho(r)\varphi(r) dV, \qquad (8)$$

where $v(r)$, $\rho(r)$ and $\varphi(r)$ are the local electron speed, charge density and electric potential, respectively, in a plasmon wave within a NC. The electron mass is denoted by $m_e$ and $n_0(r)$ is the equilibrium density of electrons in a NC. In the field theory of plasmons,[45] the above variables become quantized and the expression for the energy (8) is substituted for the Hamiltonian:

$$E_{plasmon} \to \hat{H}_{plasmon} = \int n_0(r) \frac{\hat{p}^2(r)}{2m_e} dV + \frac{1}{2}\int \hat{\rho}(r)\hat{\varphi}(r) dV.$$

This Hamiltonian involves the field operators, including the linear momentum operator, $\hat{p}(r)$. In the classical limit, the plasmonic energy (eq 8) can be also written via the local dielectric function $\varepsilon(r,\omega)$ [61]

$$E_{plasmon} = \frac{1}{4\pi}\int dV \frac{d\varepsilon(r,\omega)\cdot\omega}{d\omega} \mathbf{E}_\omega(r)\cdot\mathbf{E}_\omega^*(r), \qquad (9)$$

where $\mathbf{E}_\omega(r)$ is the amplitude of the electric field created by the surface charges of a NC. In the Drude model, the plasmonic energy stored in a spherical NP is given by the following simple equation (for the derivation, see Supporting Information):

$$E_{plasmon} = \frac{1}{2\pi} V_{NP} \cdot E_{\omega,in}^2 \left(2\varepsilon_0 + \varepsilon_{b,Drude}\right), \qquad (10)$$

where $E_{\omega,in}$ is the amplitude of the electric field inside the NC, $V_{NP}$ is the NC volume, and $\varepsilon_0$ and $\varepsilon_{b,Drude}$ are the dielectric constant of the matrix and the background dielectric constant of the metal in the Drude limit, respectively.

In this paper, we will see that the surface-scattering mechanism should be substituted by a more general hot-electron mechanism, i.e.

$$\gamma_{surf} \rightarrow \gamma_{hot-electrons}.$$

We demonstrate this generalization of the formalism using two models, dimers with narrow gaps and nanocubes with sharp vertices. We show that $\gamma_{hot-electrons}$ includes two types of hot-electron generation: (1) near the surfaces and (2) in hot-spot regions of a NC. We will see that the surface mechanism of scattering does not suffice to describe the processes in NCs with sharp tips. Moreover, the rate $\gamma_{surf}$ is also strongly affected by hot spots (see the section on hot spots). Therefore, eqs 4 and 5 can be rewritten as

$$\gamma_{plasmon} = \gamma_{Drude} + \gamma_{inter\text{-}band} + \gamma_{hot-electrons}, \qquad (11)$$

where $\gamma_{hot-electrons}$ is responsible for the generation of energetic electron-hole pairs in a NC. In NCs with hot spots, this process may occur not only near the surface, but also in hot-spot regions in the bulk. The source of these processes is the breaking of the momentum conservation for electrons at the surfaces or in electromagnetic hot-spot regions. Thus, the rate $\gamma_{hot-electrons}$ may not be always regarded as a rate of generation near the surface. One example is the excitation of hot-electrons in nanocubes with sharp vertices, where the surface curvature is too large and the surface mechanism of hot electron generation is not applicable.

The structure of the relaxation rate of a plasmon (eq 11) can be also understood from the point of view of energy conservation in a NC:

$$\frac{dE_{plasmons}}{dt} = Q_{tot} - \left(\gamma_{Drude} + \gamma_{inter-band} + \gamma_{hot-elecrons}\right) \cdot E_{plasmons} \qquad (12)$$
$$Q_{tot} = \langle \mathbf{j} \cdot \mathbf{E}_{ext} \rangle_{time}$$

where $E_{plasmons}$ is the plasmonic energy stored in a NC and $Q_{tot}$ is the rate of absorption of the external radiation. In an optically-excited NC, one can distinguish three types of dissipation of light energy:

$$Q_{total} = Q_{Drude} + Q_{inter-band} + Q_{hot-electrons}. \qquad (13)$$

The above absorption rates come from the three types of transitions in a metal NC: (1) $Q_{Drude}$ is due to the intra-band absorption, that can be described classically via a local frictional force within the Drude model; (2) $Q_{inter-band}$ is the energy spent for the generation of interband electron-hole pairs; (3) $Q_{hot-electrons}$ is the power directed to the creation of hot electrons in the sp-band. The terms $Q_{Drude}$ and $Q_{inter\text{-}band}$ are described by the classical electromagnetic theory (see the derivations below).

In the regime of CW illumination, we have $dE_{plasmons}/dt = 0$. Then, eqs 12 and 13 tell us that the plasmon relaxation mechanisms originate from the corresponding dissipations:

$$\gamma_{Drude} + \gamma_{inter\text{-}band} + \gamma_{hot-elecrons} = \frac{Q_{tot}}{E_{plasmons}} = \frac{Q_{Drude} + Q_{inter\text{-}band} + Q_{hot-electrons}}{E_{plasmons}}. \qquad (14)$$

In particular, we see that for the hot-electron mechanism:

$$\gamma_{hot-elecrons} = \frac{Q_{hot-electrons}}{E_{plasmons}}. \qquad (15)$$

On the other hand, the energy dissipation due to the generation of hot electrons is also connected to their generation rate (see the derivations above),

$$Q_{hot-electrons} \approx \hbar\omega \cdot Rate_{high-energy}. \qquad (16)$$

Then, we can combine (15) and (16) to obtain

$$\gamma_{hot-electrons} \approx \frac{\hbar\omega \cdot Rate_{high-energy}}{E_{plasmon}}, \qquad (17)$$

recovering eq 7 with $\gamma_{surf} \to \gamma_{hot-electrons}$.

This analysis shows that hot electron generation and surface scattering are intrinsically related. However, the observation of plasmon broadening induced by surface scattering is not easy for the majority of NCs, since they are typically of relatively large sizes. In fact, the rate $\gamma_{surf}$ can be identified from the absorption spectra only for relatively small NP sizes.[53,54] Simultaneously, there are other types of experiments that can provide evidence of hot electrons - photochemistry, photocurrent and time-resolved studies. These types of measurements are sensitive to the generation of energetic electrons near the NC surfaces, which are then able to transit to neighboring materials.

The efficiency of hot electron production, or the quantum parameter of the plasmon, is defined as

$$QP_{plamson} = \frac{Q_{hot-electrons}}{Q_{tot}} = \frac{Rate_{hot-electrons}}{Rate_{abs,photons}}, \qquad (18)$$

where $Q_{tot}$ is the total absorption in a NC that includes the three contributions in eq 14; $Rate_{abs,photons} = Q_{tot}/\hbar\omega$ is the rate of absorption of photons. This parameter shows how much optical energy is spent exciting energetic electrons. The efficiency of generation of hot electrons in NCs can be analytically estimated as

$$QP_{plasmon} = \frac{1}{1+\frac{a_0}{const'\cdot l_{mfp}}} \qquad (19)$$

For large sizes ($a_0 > l_{mfp}$), this expression is converted to

$$QP_{plasmon} \to const'\cdot \frac{l_{mfp}}{a_0}$$

Therefore, the power efficiency (eq 18) depends directly on the ratio $\frac{l_{mfp}}{a_0}$, whereas the ratio of the rates, $Rate_{high-energy} / Rate_{low-energy,Drude}$ (eq 3), has a more complex dependence on the NC's size. We note that eq 19 is exact in the absence of inter-band transitions. When we include the interband transitions, the mean free path in eq 19 should be changed to a shorter length, $l_{eff}$ (see a later section). Below we will demonstrate these properties both analytically and numerically.

Along with computer simulations, this paper contains simple, useful analytical equations for the generation rates of hot carriers. In particular, the rate of generation of hot carriers at the surface of NCs with simple shapes is given by the integral (see ref [46] and Supporting Information):

$$Rate_{high-energy} \approx \frac{2}{\pi^2} \times \frac{e^2 E_F^2}{\hbar} \frac{1}{(\hbar\omega)^3} \int_{S_{NC}} |E_{normal}(\theta,\varphi)|^2 \, ds, \quad (20)$$

where $E_{normal}$ is the component of the electric field normal to the NC's surface, inside the NC and the integral is taken over the whole surface; $E_F$ is the Fermi energy of the metal. An important quantum multiplier, $(\hbar\omega)^{-3}$, originates from the summation over all quantum transitions near the surfaces of a NC. Furthermore, eq 20 can be simplified for the case of a small spherical geometry:

$$Rate_{high-energy,sphere} \approx \frac{2}{\pi^2} \times \frac{e^2 E_F^2}{\hbar} \frac{1}{(\hbar\omega)^3} \frac{4\pi}{3} R_0^2 \left| \frac{3\varepsilon_0}{2\varepsilon_0 + \varepsilon_{metal}} \right| \frac{2\pi}{c_0\sqrt{\varepsilon_0}} I_0,$$

where $I_0$ is the intensity of incident light, $R_0$ is the NP radius, and $\varepsilon_0$ and $\varepsilon_{metal}$ are the dielectric constants of matrix and metal, respectively. If we need to compute the number of carriers above certain energy barrier, we can use again eq 20, but with an additional factor $(\hbar\omega - \Delta E_b)/\hbar\omega$:

$$Rate_{high-energy, \varepsilon > \Delta E_b} = \frac{2}{\pi^2} \times \frac{e^2 E_F^2}{\hbar} \frac{\hbar\omega - \Delta E_b}{(\hbar\omega)^4} \int_{S_{NC}} |E_{normal}(\theta, \varphi)|^2 \, ds,$$

where $\Delta E_b$ is the barrier height. Figure 1d shows the excited electrons that can propagate above the barrier into the contact as a red area.

## Quantum formalism for the effect of generation of hot electrons.

In this section we describe the formalism used to compute the spectral rates of generation and the electron energy distribution functions in optically-driven NCs. A convenient quantum formalism to derive the distributions of hot electrons is the one-particle density matrix $\rho_{nm}(t)$. Below we list the main results for the problem of hot electron generation, more details are given in the Supporting Information. The main derivations for the formalism used in this text have been reported in our previous papers.[28,29,40,42] In the present one we make some important developments and generalizations for our previous quantum formalism. In particular, we formulate here the problem of hot electrons in terms of the generation rates and the detailed-balance equation, whereas our previous calculations were mostly done for the non-equilibrium populations of electrons.

The one-particle density matrix contains the information about the statistical distribution of excited carriers in a plasmonic wave:[62]

$$\rho_{nm}(t) = \langle \Psi(t) | \hat{c}_m^\dagger \hat{c}_n | \Psi(t) \rangle. \qquad (21)$$

Here $|\Psi(t)\rangle$ is the time-dependent many-body function of the electron system under illumination, and $\hat{c}_n$ and $\hat{c}_m^\dagger$ are the second quantization operators for the corresponding single-particle states. The kinetic equation of motion for the electronic density matrix reads[62]

$$\hbar\frac{\partial\hat{\rho}}{\partial t} = i\left[\hat{\rho},\hat{H}_{sp}\right] - \hat{\Gamma}(\hat{\rho}), \qquad (22)$$

where $\hat{\rho}$ is the density matrix operator used to generate the matrix elements, $\rho_{nm} = \langle n|\hat{\rho}|m\rangle$; the operator $\hat{\Gamma}$ describes the relaxation of the electron system and $\hat{H}_{sp}$ is the single-particle Hamiltonian, which should be written within the random-phase approximation (RPA). A single-particle quantum state $|n\rangle$ includes both orbital and spin quantum numbers, $|n\rangle = |n_{orb},s\rangle$, where $s = \pm 1/2$. The expression in eq 22 naturally leads to the equation of energy conservation given in eq 12. The single-particle Hamiltonian in eq 22 has the form:

$$\hat{H}_{sp}(\mathbf{r},t) = -\frac{\hbar^2}{2m_0}\nabla^2 + U_0(\mathbf{r}) + \hat{V}_{optical}(\mathbf{r},t), \qquad (23)$$

where $U_0(\mathbf{r})$ is the confining potential of a NC and $\hat{V}_{optical}(\mathbf{r},t)$ is the time-dependent optical perturbation:

$$\hat{V}_{optical}(\mathbf{r},t) = V_a(\mathbf{r})e^{-i\omega t} + V_b(\mathbf{r})e^{i\omega t}$$
$$V_a = e\varphi_\omega(\mathbf{r}), \quad V_b = e\varphi_\omega^*(\mathbf{r}).$$

The time-dependent perturbation $\hat{V}_{optical}(\mathbf{r},t)$ is a self-consistent potential that includes both the external field and the fields of induced plasmonic changes. Then, we take the relaxation operator in the following form:

$$\langle m|\hat{\Gamma}\hat{\rho}|n\rangle = \begin{cases} \langle n|\hat{\Gamma}\hat{\rho}|n\rangle, & m = n \\ \hbar\dfrac{\rho_{mn}}{\tau_p}, & m \neq n \end{cases}. \qquad (24)$$

Here the off-diagonal part assumes fast dephasing due to quasi-elastic scattering in the momentum space, whereas the diagonal part can be more complex (see next section). For the off-diagonal part, we define the momentum relaxation rate in the energy units, $\gamma_p = \hbar/\tau_p$ (Table 1). These rates will be taken from the Drude terms of the empirical dielectric functions[63] (see Supporting Information).

Under the steady-state optical excitation, the system has a specific dynamical balance for each quantum state. In other words, the occupations of the states do not change with time and we obtain from the master equation (eq 22):

$$\begin{aligned}\frac{\partial \rho_{nn}}{\partial t} &= G_n - R_n = 0, \\ G_n &= \frac{i}{\hbar}\langle n|[\hat{\rho}, \hat{H}_{sp}]|n\rangle, \\ R_n &= \frac{1}{\hbar}\langle n|\hat{\Gamma}(\hat{\rho})|n\rangle,\end{aligned} \qquad (25)$$

where $G_n$ and $R_n$ are the rates of generation and relaxation for the quantum state $|n\rangle$. Under the steady-state conditions, these two rates are, of course, the same. Then, we can look at the rate of carrier generation in the system (see eq 2):

$$Rate_{e,tot} = \frac{dN}{dt} = \sum_n G_n, \qquad (26)$$

Correspondingly, the optical-energy dissipation takes the form:

$$Q_{tot,quantum} = \sum_n \varepsilon_n G_n = \sum_n \varepsilon_n R_n. \qquad (27)$$

The above eqs 26 and 27 are also consistent with the dynamic equation of energy conservation. The solution for eq 25 within the time-dependent perturbation theory was found and described in details in refs [29,40]. The result for the hot-electron generation rates taken from refs [29,40] has the form:

$$G_n = \frac{2}{\hbar} \sum_{n'} (f_{n'} - f_n) \left[ |V_{nn',a}|^2 \frac{\gamma_p}{(\hbar\omega - \varepsilon_n + \varepsilon_{n'})^2 + \gamma_p^2} + |V_{nn',b}|^2 \frac{\gamma_p}{(\hbar\omega + \varepsilon_n - \varepsilon_{n'})^2 + \gamma_p^2} \right], \qquad (28)$$

where $f_n$ is the equilibrium Fermi function describing the occupation of the single-particle state $|n\rangle$. The optical matrix elements in eq 28 are given by

$$V_{nn',a} = \langle n | e\varphi_\omega(r) | n' \rangle, \quad V_{nn',b} = \langle n | e\varphi_\omega^*(r) | n' \rangle. \qquad (29)$$

Below we will give a convenient approach to compute such matrix elements using widely used electrodynamic software packages, like COMSOL. The main formulas (derived from eqs 26 and 28) to compute the rates of generation and dissipation read:

$$Rate_{e,tot} = \frac{dN}{dt} = \frac{2}{\hbar} \sum_{n,n'} (f_{n'} - f_n) \left[ |V_{nn',a}|^2 \frac{\gamma_p}{(\hbar\omega - \varepsilon_n + \varepsilon_{n'})^2 + \gamma_p^2} + |V_{nn',b}|^2 \frac{\gamma_p}{(\hbar\omega + \varepsilon_n - \varepsilon_{n'})^2 + \gamma_p^2} \right],$$

$$Q_{tot} = \frac{2}{\hbar} \sum_{n,n'} \varepsilon_n (f_{n'} - f_n) \left[ |V_{nn',a}|^2 \frac{\gamma_p}{(\hbar\omega - \varepsilon_n + \varepsilon_{n'})^2 + \gamma_p^2} + |V_{nn',b}|^2 \frac{\gamma_p}{(\hbar\omega + \varepsilon_n - \varepsilon_{n'})^2 + \gamma_p^2} \right]. \qquad (30)$$

At this point, we will also define the rates for the low-energy and high-energy electrons (Drude and hot carriers in Figure 1e):

$$Rate_{e,tot} = Rate_{low-energy} + Rate_{high-energy}$$
$$Rate_{low-energy} = \sum_{E_F < \varepsilon_n < \delta E_{crit}} G_n, \quad Rate_{high-energy} = \sum_{\varepsilon_n > \delta E_{crit}} G_n \qquad (31)$$

Here $\delta E_{crit}$ is the threshold energy that marks the boundary between low-energy and hot electrons (see next section). Then, for the optical absorption

$$Q_{tot,quantum} = Q_{low-energy} + Q_{high-energy}$$
$$Q_{low-energy} = \sum_{|\varepsilon_n - E_F| < \delta E_{crit}} \varepsilon_n G_n, \quad Q_{high-energy} = \sum_{|\varepsilon_n - E_F| > \delta E_{crit}} \varepsilon_n G_n \qquad (32)$$

To see how hot electrons are distributed over the energy range, we now introduce the energy distribution for the generation rate of excited carriers. The mathematical form of such distribution is

$$R_{e,\delta}(\varepsilon) = \frac{dN}{d\varepsilon dt} = \sum_n G_{nn} \cdot \delta(\varepsilon - \varepsilon_n), \qquad (33)$$

where $\delta(\varepsilon - \varepsilon_n)$ is the delta function. This distribution contains all the information about how electrons are excited among the manifold of excited single-particle states in a many-electron NC. However, a more practical distribution function, with a "broadened" delta function is:

$$R_e(\varepsilon) = \frac{dN}{d\varepsilon dt} = \sum_n G_n \cdot P(\varepsilon - \varepsilon_n), \qquad (34)$$

where $P(\varepsilon - \varepsilon_n)$ is defined as

$$P(\varepsilon) = \begin{cases} \dfrac{1}{\delta\varepsilon}, & |\varepsilon| < \delta\varepsilon/2 \\ 0, & |\varepsilon| > \delta\varepsilon/2 \end{cases},$$

Here $\delta\varepsilon$ is the broadening, that should be taken as a small value in the numerical computations. In what follows, we will use eq 34 for our numerical simulations.

The energy distribution for the rate in eq 34 is very useful for calculations of the integrated rates and the dissipations. To find the total rates we simply integrate:

$$Rate_{e,tot} = \int_{E_F}^{\infty} R_e(\varepsilon)d\varepsilon$$

$$Rate_{low-energy} = \int_{E_F}^{\delta E_{crit}} R_e(\varepsilon)d\varepsilon, \quad Rate_{high-energy} = \int_{\delta E_{crit}}^{\infty} R_e(\varepsilon)d\varepsilon$$

For the dissipations, we have similarly

$$Q_{tot,quantum} = \int_{0}^{\infty} \varepsilon \cdot R_e(\varepsilon)d\varepsilon$$

$$Q_{low-energy} = \int_{|\varepsilon_n - E_F| < \delta E_{crit}} \varepsilon \cdot R_e(\varepsilon)d\varepsilon, \quad Q_{high-energy} = \int_{|\varepsilon_n - E_F| > \delta E_{crit}} \varepsilon \cdot R_e(\varepsilon)d\varepsilon$$

(35)

In addition, we now define the distribution function of excited electrons in a localized plasmonic wave:

$$\delta n(\varepsilon) = \frac{dN}{d\varepsilon} = \sum_{n} \rho_{nn} \cdot P(\varepsilon - \varepsilon_n).$$

Finally, it is worthwhile to note that the above equations for the rates can be also derived from the Fermi's golden rule in which the energy broadening is taken as $\gamma_p = \gamma_{Drude}$, the broadening in the Drude part of the dielectric constant of the metal. However, it is important to make a consistent derivation of the rates from the master kinetic equation (eq 22) because this method allows us to identify the broadening of the states in the single-particle rates (28).

Table 1: Parameters used in the calculations of the hot electron distributions. Note that the rates and plasmon frequency are given here in energy units.

| Parameter | Au | Ag |
|---|---|---|
| $\gamma_p$ | 0.078 eV | 0.020 eV |
| $\hbar/\tau_{\varepsilon,phonons}$ | 0.0013 (0.5 ps) | 0.0013 eV |
| $\omega_{p,Drude}$ | 9.1 eV | 9.3 eV |
| Work Function | 4.6 eV | 4.7 eV |
| Fermi Energy | 5.5 eV | 5.76 eV |
| $\varepsilon_{b,Drude}$ | 9.07 | 7.0 |

**Electron distribution functions for plasmons.**

Under CW illumination, the populations of electronic states are in the steady state. Then, according to eq 25, the state occupations $\rho_{nn}$ should be found from the balance equations for each state:

$$G_n = R_n.$$

The generation rate $G_n$ under weak illumination was found (eq 28) and the resulting equation to be solved becomes:

$$\frac{1}{\hbar}\langle n|\hat{\Gamma}(\hat{\rho})|n\rangle = G_n. \qquad (36)$$

Clearly the occupations $\rho_{nn}$ depend on the choice of the relaxation operator. We now consider two cases for this operator. The first approach allows us to obtain a simple and fast solution for the stationary distribution of state probabilities, and the second case is more complex, but can be solved either numerically or semi-quantitatively.

**Approach 1.** The relaxation operator can be taken to involve a single energy-relaxation time:

$$\frac{1}{\hbar}\langle n|\hat{\Gamma}\hat{\rho}|n\rangle = \frac{\hat{\rho}_{nn} - f_F(\varepsilon_n, T_{lattice})}{\tau_\varepsilon}, \qquad (37)$$

where $\tau_\varepsilon$ is the characteristic energy relaxation time that can be taken from experiments; the corresponding rate is defined as $\gamma_\varepsilon = \hbar/\tau_\varepsilon$. The function $f_F(\varepsilon_n, T_{lattice})$ is the Fermi distribution at the lattice temperature. This approach was used by us in refs [28,29,40–42] and the relaxation time in this case should be considered as phonon-mediated. This approach has two advantages: (1) it is simple and (2) it introduces a second relaxation time ($\tau_\varepsilon$) that is typically much longer that the momentum relaxation time ($\tau_p$). Table 1 lists the phenomenological relaxation times taken from the experiments and we see that $\tau_\varepsilon \gg \tau_p$ or $\gamma_\varepsilon \ll \gamma_p$; in particular, the energy relaxation time was taken from ref 64. This approach sets up two very different time scales for relaxation that we have typically see in experiments. In this simple approach, the solution for the distribution function is easy to find from eqs 28, 36 and 37:

$$\delta\rho_{nn} = \rho_{nn} - f_F(\varepsilon_n, T_{lattice}) =$$
$$= \frac{1}{\gamma_\varepsilon}\frac{2}{\hbar}\sum_{n'}(f_{n'} - f_n)\left[|V_{nn',a}|^2 \frac{\gamma_p}{(\hbar\omega - \varepsilon_n + \varepsilon_{n'})^2 + \gamma_p^2} + |V_{nn',b}|^2 \frac{\gamma_p}{(\hbar\omega + \varepsilon_n - \varepsilon_{n'})^2 + \gamma_p^2}\right] \qquad (38)$$

Then, we can calculate numerically the distribution of excited electrons over energy as:

$$\delta n(\varepsilon) = \frac{dN}{dE} = \sum_{\mathbf{n}} \rho_{nn} \cdot P(\varepsilon - \varepsilon_n).$$

**Approach 2.** The second approach is more advanced and involves two different energy-relaxation mechanisms: photon emission and electron-electron scattering:

$$R_n = \frac{1}{\hbar}\langle n|\hat{\Gamma}\hat{\rho}|n\rangle = \frac{\rho_{nn} - f_F(\varepsilon_n, T_{lattice})}{\tau_{\varepsilon, phonons}} + R_{ee}$$

$$R_{ee} = \sum_{n' \neq n} w^{e-e}_{n',n} \cdot \rho_{nn}(1-\rho_{n'n'}) - w^{e-e}_{n,n'} \cdot \rho_{n'n'}(1-\rho_{nn})$$

(39)

where $R_{ee}$ is the electron-electron scattering operator, in which $w^{e-e}_{n',n}$ are the rates for the electron-electron scattering processes which lead to relaxation of an excited electron via creation of electron-hole pairs in the Fermi sea (Figure 2b). The operator $R_{ee}$ in eq 39 is taken in the form proposed in ref 38 and 59. Eq 39 cannot be solved analytically for the function $\rho_{nn}$ and certainly needs a numerical approach. For example, it was recently solved numerically in ref [39]. We can find, however, a convenient semi-quantitative approximation for the solution of eq 39. For this, we consider now the e-e relaxation time of a single electron. One convenient approximation for this parameter reads:[66,67]

$$\tau_{\varepsilon,ee} = \tau_0 \frac{E_F^2}{(\varepsilon - E_F)^2}, \quad (40)$$

where the parameter $\tau_0 = 6\ fs$ for gold. The e-e time has the characteristic behavior $(\varepsilon - E_F)^{-2}$ that was first discovered as a part of the Fermi liquid theory.[68] In the text below we will show the calculated characteristic electron-electron scattering time $\tau_{\varepsilon,e-e}$, together with typical value of the phonon-scattering time $\tau_{\varepsilon,phonons}$ (the corresponding rates will be defined in this paper as $\gamma_{\varepsilon,e-e} = \hbar/\tau_{\varepsilon,e-e}$ and $\gamma_{\varepsilon,phonons} = \hbar/\tau_{\varepsilon,phonons}$). The typical physical situation in metals is the following: the e-e scattering mechanism dominates at high energies, whereas, for small excess energies $\varepsilon - E_F$, the phonon mechanism should be applied. In fact, the e-e relaxation mechanism typically leads to rather short relaxation times (in the fs range!) for high-energy optically-excited carriers (see Figure 7 below).

Another important property of the e-e scattering is that the electron-electron operator in eq 39 conserves the electronic energy:

$$\sum_n \varepsilon_n R_{ee,n} = 0. \qquad (41)$$

Therefore, we can use the following semi-quantitative approach to the relaxation operator:

$$\begin{aligned}
R_{nn} &\approx \left( \frac{\rho_{nn}}{\tau_{\varepsilon,ee}} + \frac{\rho_{nn}}{\tau_{\varepsilon,phonons}} \right), \quad |\varepsilon - E_F| > \delta E_{crit} \\
R_{nn} &\approx \frac{\rho_{nn} - f_F(\varepsilon_n, T_{lattice})}{\tau_{\varepsilon,phonons}} - F_{gen,e-e}(\varepsilon), \quad |\varepsilon - E_F| < \delta E_{crit} \\
F_{gen,e-e}(\varepsilon) &= \left( \frac{Q_{eff}}{3.3 \cdot k_B T_{eff} \cdot DOS_F} \right) \frac{1}{k_B} \frac{\partial f_F(\varepsilon, T_{eff})}{\partial T} \\
Q_{eff} &= \sum_{|\varepsilon_n - E_F| > \delta E_{crit}} \varepsilon_n \cdot G_n \left( \frac{\gamma_{\varepsilon,ee}(\varepsilon_n)}{\gamma_{\varepsilon,phonons} + \gamma_{\varepsilon,ee}(\varepsilon_n)} \right)
\end{aligned} \qquad (42)$$

In eq 42, the parameter $k_B T_{eff}$ is the characteristic width of the energy distribution for the relaxed electrons, $\delta E_{crit}$ is the threshold energy for the population of high-energy electrons (this quantity is described in more detail in a later section and also in Supporting Information), and $DOS_F$ is the density of states at the Fermi level. The effective temperature $k_B T_{eff}$ for the relaxed electrons should be obtained from exact calculations using the relaxation operator (eq 39). Describing the steady-state non-equilibrium electron distributions, we will choose the following values for this parameter: $k_B T_{eff} = 0.1\,eV$ and $0.04\,eV$, for Au NCs with sizes of 4 *nm* and 24 *nm*. Despite some uncertainties in the parameters, this simplified approach will allow us to reconstruct the physical picture of the steady-state hot-electron distributions (see Figure 7 that can be found in the next numerical section). Using the above approximations, we can now easily obtain the occupations from eqs 41 and 42:

$$\delta\rho_{nn} = \tau_{\varepsilon,\text{eff}} G_n, \quad |\varepsilon - E_F| > \delta E_{crit};$$

$$\delta\rho_{nn} = \tau_{\varepsilon,\text{phonons}} \left[ G_n + \left( \frac{Q_{eff}}{3.3 \cdot k_B T \cdot DOS_F} \right) \frac{1}{k_B} \frac{\partial f_F(\varepsilon)}{\partial T} \right], \quad |\varepsilon - E_F| < \delta E_{crit}, \quad (43)$$

$$\tau_{\varepsilon,\text{eff}} = \frac{\tau_{\varepsilon,ee} \tau_{\varepsilon,\text{phonons}}}{\tau_{\varepsilon,ee} + \tau_{\varepsilon,\text{phonons}}}$$

For the energy distribution of excited electrons, we now obtain:

$$\delta n(\varepsilon) = \sum_n \tau_{\varepsilon,\text{eff}} G_n, \quad |\varepsilon - E_F| > \delta E_{crit}$$

$$\delta n(\varepsilon) = \tau_{\varepsilon,\text{phonons}} \left[ \sum_n G_n + \left( \frac{Q_{eff}}{3.3 \cdot k_B T} \right) \frac{1}{k_B} \frac{\partial f_F(\varepsilon)}{\partial T} \right], \quad |\varepsilon - E_F| < \delta E_{crit} \quad (44)$$

The functions in eq 44 describe two important physical properties of the steady-state distribution of excited electron in a plasmon wave. (1) These equations conserve the total electronic energy during the e-e relaxation process and (2) these distributions reflect the dramatic shortening of the electronic lifetime for large electron energies owing to the e-e collisions. In a later section, we will apply this approach to the case of Au NCs and see strong effects coming from the e-e collisions.

**Mathematical approaches to computation of hot electrons in small and large NCs.**

In this section we describe how to reliably compute the optical matrix elements, generation rates and nonequilibrium distribution functions of plasmonic NCs. Since we work with monochromatic fields, it is convenient to define the electric fields via their complex amplitudes. The external field is taken as

$$\mathbf{E}_{external} = 2 \cdot \mathbf{E}_0 \cos(\omega \cdot t) = \mathbf{E}_0 \cdot e^{-i\omega t} + \mathbf{E}_0 \cdot e^{+i\omega t}.$$

Here $\mathbf{E}_0$ is the field amplitude. Using this input, we compute numerically the near-field, which may have a complex spatial structure, using the COMSOL software package. The field inside a NC is then $\mathbf{E}_{optical} = \mathbf{E}_\omega \cdot e^{-i\omega t} + \mathbf{E}_\omega^* \cdot e^{+i\omega t}$, where the complex amplitude $\mathbf{E}_\omega$ is evaluated numerically.

The optical matrix elements (eq 29) involve the total potential $\varphi_\omega(\mathbf{r})$. However, it is more convenient to express it via the local electric field since typically numerical programs (such as

COMSOL, Lumerical, etc.) directly compute these fields. Therefore, we now rewrite eq 29 in the following form:

$$\varphi_{nn',a} = \langle n|\varphi_\omega(r)|n'\rangle = \frac{\hbar^2}{m(\varepsilon_n - \varepsilon_{n'})} \int dV \cdot \psi_n \left(\mathbf{E}_\omega \cdot \vec{\nabla} \psi_{n'}\right),$$

where $\mathbf{E}_\omega = -\vec{\nabla}\varphi_\omega$ is the complex amplitude of the total electric field inside the nanocrystal. To see the strength and distribution of the near-fields in a NC, we define the local enhancement factor:

$$P_{local}(\mathbf{r}) = \frac{\mathbf{E}_\omega \cdot \mathbf{E}_\omega^*}{E_0^2}.$$

To appreciate the total field enhancement inside and at the surface of the NC, one can evaluate the following integrals:

$$P_V(\omega) = \frac{1}{V_{NP}} \int_V \frac{\mathbf{E}_\omega \cdot \mathbf{E}_\omega^*}{E_0^2} dV, \quad P_S(\omega) = \frac{1}{S_{NP}} \int_S \frac{E_{\omega,norm} \cdot E_{\omega,norm}^*}{E_0^2} ds,$$

where $V_{NP}$ and $S_{NP}$ are the NC's volume and surface area, respectively; $E_{\omega,norm}$ is the electric field normal to the surface. Then, the two components in the dissipation expression can be evaluated via the bulk dielectric constant:

$$\begin{aligned}
Q_{classical} &= Q_{Drude} + Q_{inter\text{-}band} = \left\langle \int_{NC} dV \, \mathbf{j} \cdot \mathbf{E} \right\rangle_{time} = \text{Im}(\varepsilon_{metal}) \frac{\omega}{2\pi} \int_{NC} dV \, \mathbf{E}_\omega \cdot \mathbf{E}_\omega^* \\
Q_{Drude} &= \text{Im}(\Delta\varepsilon_{Drude}) \frac{\omega}{2\pi} \int_{NC} dV \, \mathbf{E}_\omega \cdot \mathbf{E}_\omega^* \\
Q_{interband} &= \text{Im}(\varepsilon_{inter\text{-}band}) \frac{\omega}{2\pi} \int_{NC} dV \, \mathbf{E}_\omega \cdot \mathbf{E}_\omega^*
\end{aligned} \quad (45)$$

In the above equations we have used the following property of the bulk dielectric constant of a metal:

$$\varepsilon_{metal} = \varepsilon_{inter\text{-}band} + \Delta\varepsilon_{Drude},$$

$$\Delta\varepsilon_{Drude}(\omega) = \varepsilon_{intra\text{-}band}(\omega) = -\frac{\omega_p^2}{\omega(\omega+i\gamma_p)},$$

$$\varepsilon_{inter\text{-}band}(\omega) = \varepsilon_{metal}(\omega) + \frac{\omega_p^2}{\omega(\omega+i\gamma_p)}.$$

Here $\varepsilon_{metal}$ is the empirical dielectric constant of a bulk metal, that should be taken from tables in the literature.[63] The Drude parameters for the term $\Delta\varepsilon_{Drude}$ are given in Table 1 and were evaluated from fitting to the empirical dielectric constant in the long wavelength limit. In the long wavelength intervals, we can neglect the interband transitions and use the Drude dielectric function,

$$\varepsilon_{metal,Drude}(\omega) = \varepsilon_{b,Drude} - \frac{\omega_p^2}{\omega\cdot(\omega+i\gamma_{Drude})},$$

where $\varepsilon_{b,Drude}$ is the long-wavelength background dielectric constant given in Table 1. This constant is typically found from the fit between Drude and experimental dielectric constants. Finally, the following relations are useful for optical calculations:

$$\sigma_{abs}(\omega) = \frac{Q_{classical}}{I_0}, \quad I_0 = \frac{c_0\sqrt{\varepsilon_0}}{2\pi}\cdot E_0^2,$$

where $\sigma_{abs}$ and $I_0$ are the cross section and the intensity of incident light, respectively.

An important element of the calculations for the electronic spectral functions is the averaging over the NC size. [28,29,40–42] In the following sections, we show examples of how strongly this averaging can influence numerical data. The nonequilibrium energy distributions of the NPs typically oscillate strongly as a function of energy due to the discreteness of the single-electron energy levels. Physically, this averaging arises since NCs in a solution typically have a dispersion of sizes. For spherical particles the numerical results are averaged according to:

$$\overline{f}(\varepsilon) = \frac{1}{\sqrt{\pi}\delta_R} \int_{-\infty}^{+\infty} dR \cdot e^{-\left(\frac{R-R_0}{\delta_R}\right)^2} \cdot f(\varepsilon, R), \qquad (46)$$

where $f(\varepsilon)$ is a spectral function ([$R_e(\varepsilon)$, $\delta n(\varepsilon)$, $Rate_{e,tot}$, etc.). The parameter $\delta_R$ is the broadening describing the dispersion of NC sizes and $R_0$ is the average radius of the ensemble of spherical NCs. The typical value used for the broadening parameter is $\delta_R = 1\,nm$. For the smallest NCs, we adopted $\delta_R = 0.45\,nm$. For the cubes, we apply the approach of eq 46 to the nanocube size, $L$, using a size dispersion parameter $\delta_L = 1\,nm$. Also, in the process of averaging, we keep the rounding of the corners of the nanocubes at $L/50$. In the paper, we explore the case of very sharp nanocube edges and vertices to demonstrate qualitatively the effect of hot spots.

## Nanospheres made of Drude metal, Au and Ag.

Now we apply our formalism to a simple spherical geometry. Along with numerical results, we will describe simple and efficient analytical approaches to compute the hot electron distributions in simple geometries. As an important physical model system we first compute the case of a NC made of a simple metal (Drude metal) with only one parabolic band. In the next step, we will give results for the intraband generation of carriers in real metals, gold and silver. In particular, we will see that the choice of material system can be crucial. Silver NCs generate hot electrons more efficiently than gold ones since silver has a longer electron mean free path.

For spherical particles in the quasi-static limit, the internal electric field has a simple analytical form,

$$\mathbf{E}_{\omega,in} = \mathbf{E}_0 \frac{3\varepsilon_0}{\varepsilon_{metal}(\omega) + 2\varepsilon_0},$$

where $\varepsilon_0$ is the dielectric constant of the matrix; in the following calculations we will take $\varepsilon_0 = 2$. Correspondingly, the dissipation rates and the enhancement factors become:

$$Q_{Drude} = \frac{\omega}{2\pi} \cdot V_{NC} \left|\mathbf{E}_{\omega,in}\right|^2 \cdot \mathrm{Im}(\Delta\varepsilon_{Drude})$$

$$Q_{inter\text{-}band} = \frac{\omega}{2\pi} \cdot V_{NC} \left|\mathbf{E}_{\omega,in}\right|^2 \cdot \mathrm{Im}(\varepsilon_{inter\text{-}band})$$

$$P_V(\omega) = \left|\frac{3\varepsilon_0}{\varepsilon_{metal} + 2\varepsilon_0}\right|^2, \quad P_S(\omega) = \frac{1}{3}\left|\frac{3\varepsilon_0}{\varepsilon_{metal} + 2\varepsilon_0}\right|^2$$

where $V_{NC} = \frac{4\pi R_0^3}{3}$. The single-electron states of a spherical well have the standard form:

$$\psi_n(r,\theta,\phi) = R_l\left(\frac{\alpha_{n_r l} r}{R}\right) \cdot Y_{lm}(\theta,\phi),$$

$$\varepsilon_n = \frac{\hbar^2 \alpha_{n_r l}^2}{2m_0 R^2},$$

where $R_l(r)$ are the Spherical Bessel functions, $\alpha_{n_r l}$ are the radial coefficients, and $Y_{lm}(\theta,\phi)$ are the spherical harmonics. The state $|n\rangle$ is thus characterized by three quantum numbers, $n_r, l, m$. These wave functions and their energies are used in the optical matrix elements given by eq 29.

# A nanosphere made of Drude metal: The effects of the size and momentum-relaxation time

$\hbar\omega = 2.36\ eV$

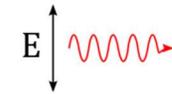
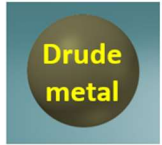

**a)**
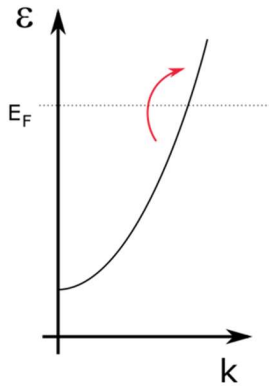

**b)**
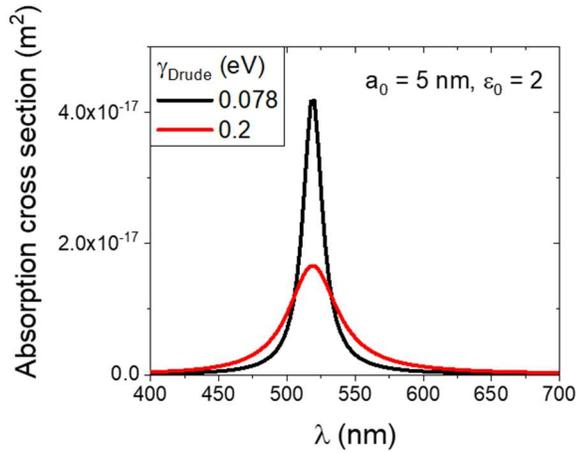

**c) Quantum limit of the plasmon**
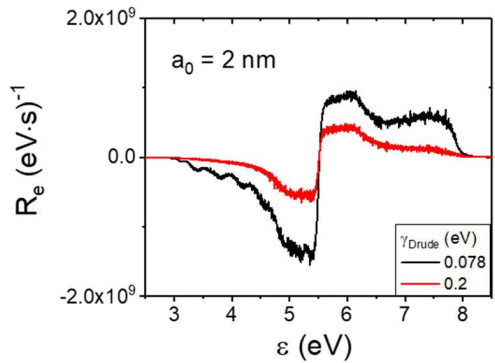

**d) Intermediate regime: Hot + Drude electrons**
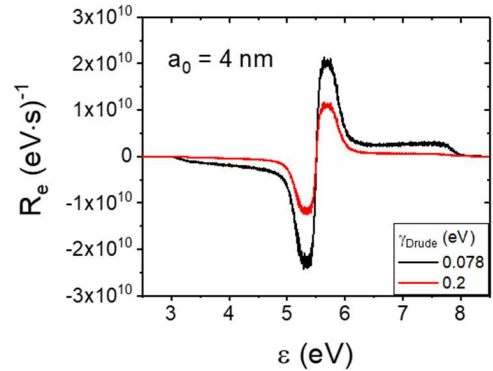

**e) Classical Drude limit**
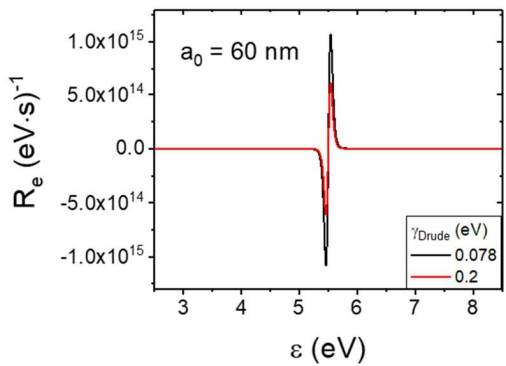
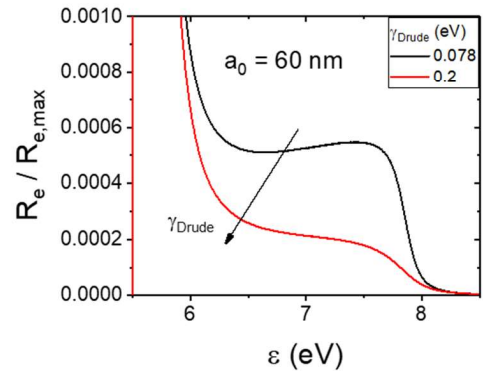

**Figure 3: (a,b)** Diagram of intraband transitionss in a Drude metal and calculated absorptions for the model with $\varepsilon_{b,Drude} = 10.5$, $\omega_P = 9.1\,eV$ and $\gamma_{Drude} = 0.078\,eV, 0.2\,eV$. **(c-d)** Spectral rates of hot-electron generation for nanospheres of different sizes. Panel (e) shows the case of a large NC which corresponds to the classical limit. Plot on the right-hand side in (e) is the spectrum of hot electrons generated due to the quantum surface effects. The incident flux is $I_0 = 3.6 \cdot 10^3\,W/cm^2$.

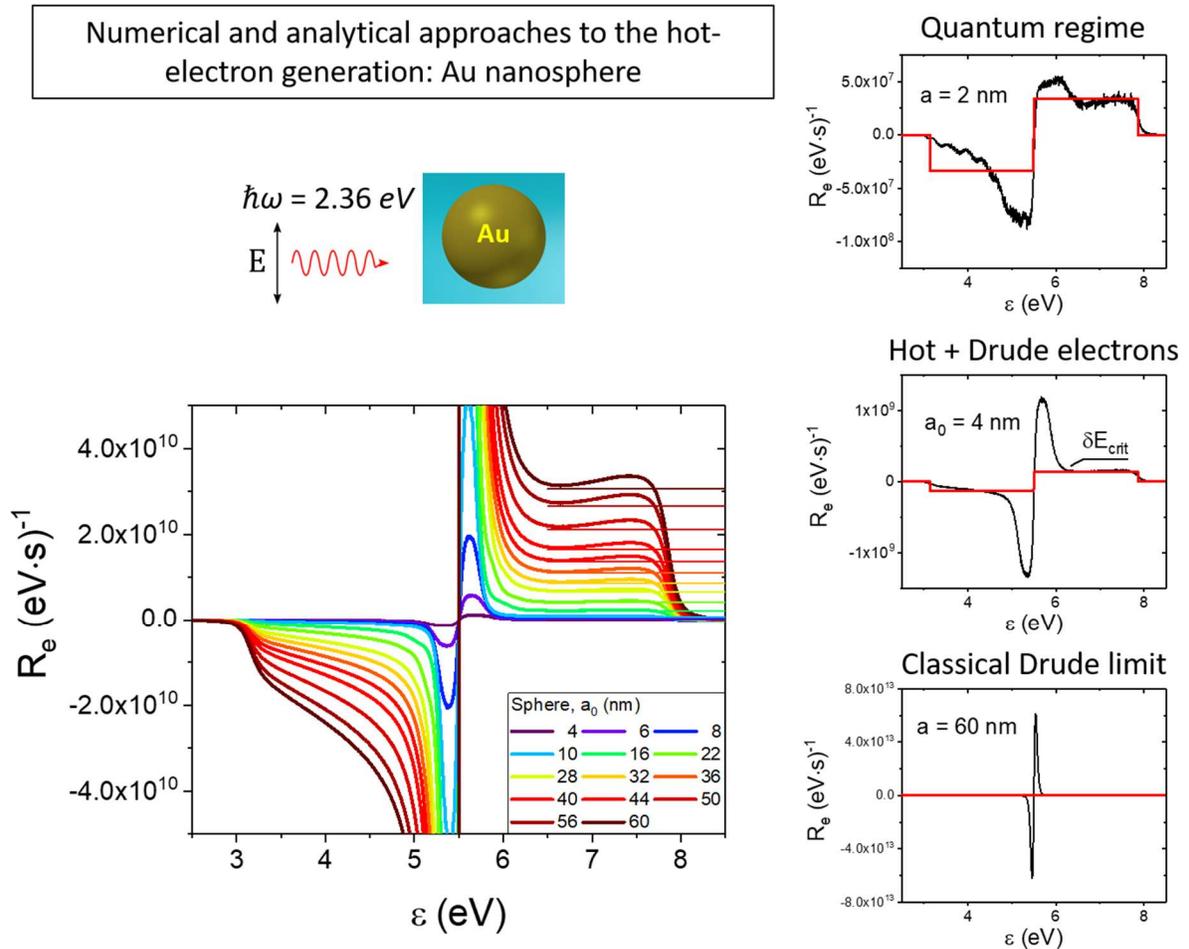

**Figure 4.** Rates of hot-electron generation for Au spheres of different sizes. On the right panels, we show the rates for three selected sizes, accompanied by analytical results as red lines. The horizontal lines in all four panels show the analytical results (eq 49) for the rates at the plateaus.

We see how the quantum plasmonic state in a small NC evolves towards the classical Drude-like plasmon. The incident flux is $I_0 = 3.6 \cdot 10^3 W/cm^2$.

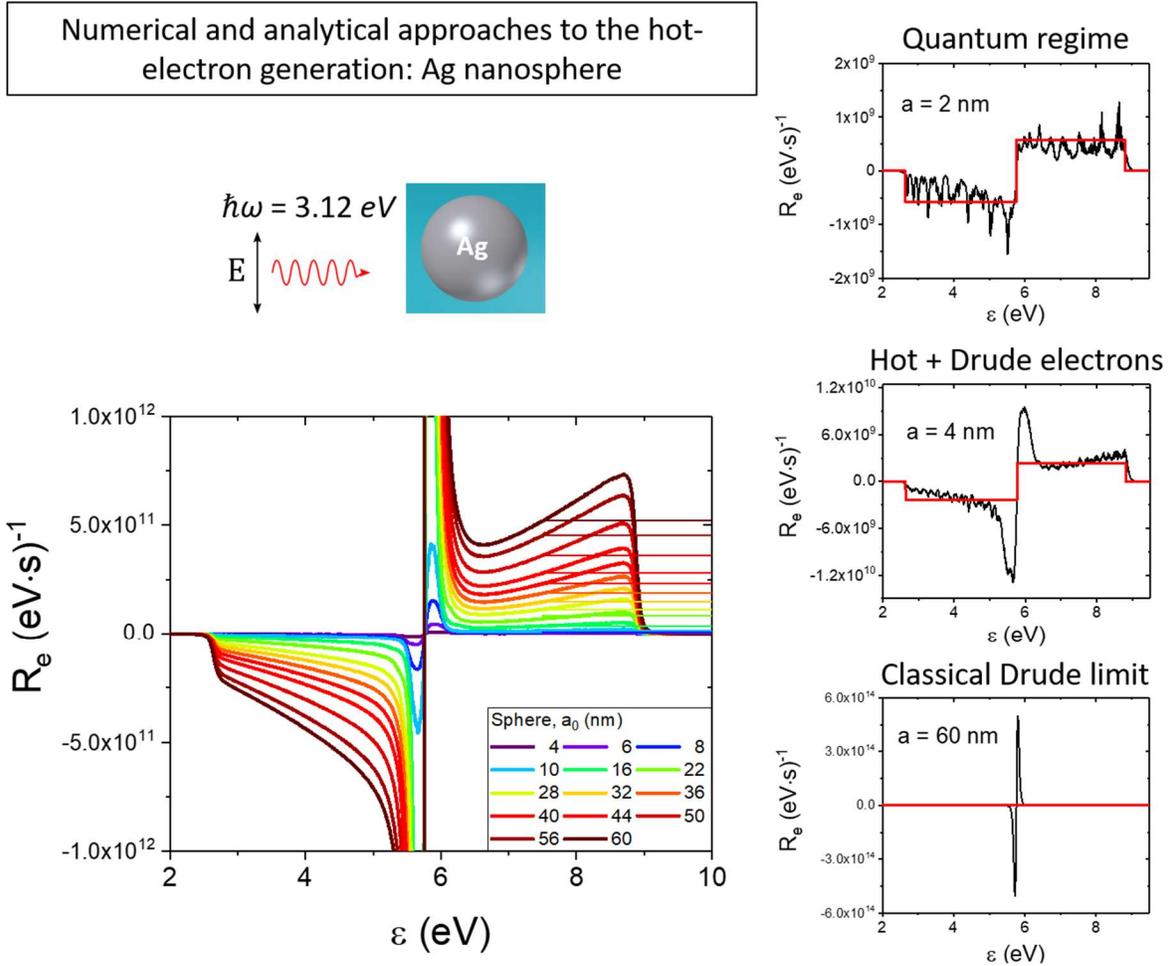

**Figure 5.** Rates of hot-electron generation for Ag spheres of different sizes. On the right panels, we show the rates for three selected sizes, accompanied by analytical results as red lines. The horizontal lines in all four panels show the analytical results (eq 49) for the rates at the plateaus. The electron distribution in small NCs is governed by the quantum transitions, whereas large NCs

have an electron distribution corresponding to the classical Drude-like plasmon. The incident flux is $I_0 = 3.6 \cdot 10^3 W/cm^2$.

**Computational data for the hot-electron rates.** Figures 3, 4 and 5 show the results for the intraband transitions in three material systems, ideal Drude metal, gold and silver. The optical energies are taken equal to the plasmonic peak frequencies the NCs (Figures 1c, 3b). Throughout this paper we will use the following value for the incident light intensity: $I_0 = 3.6 \cdot 10^3 W/cm^2$. In these figures, we show the spectra of the rates of generation and how such spectra evolve with increasing NC size (Figures 3, 4 and 5) and with increasing relaxation rate (Figure 3). The changes in the spectra are dramatic. For large NCs, the majority of excited electrons is found in the interval with low excitation energies, $E_F - \delta E_{crit} < \varepsilon < E_F + \delta E_{crit}$, where $\delta E_{crit}$ is the critical energy. At this energy $\delta E_{crit}$, the spectral rate of generation shows a transition from the Drude peak at low excitation energies to the quantum plateau (Figures 3d, 4 and S3). This energy is characterized in detail in Figure S3. A simple physical formula for this critical energy is also given below. Such low-energy excited electrons are described by the classical Drude theory. For small NCs, we observe a very different picture (Figures 3, 4 and 5): The majority of electrons have high energies and the quantum effects dominate in the electron distribution functions in the plasmonic wave.

Along with the NC size, the momentum relaxation parameter, $\gamma_p = \hbar/\tau_p$, also plays an important role (Figure 3c). The ratio between the plateau value and maximum value of the function $R_e(\varepsilon)$ is inversely proportional to the Drude momentum relaxation rate:[29]

$$\frac{R_{e,\,at\,plateau}}{R_{e,\max(\min)}} \propto 1/\gamma_p.$$

We see this important dependence in Figure 3e (right panel) that depicts a model metal NC in the classical limit for three values for the momentum relaxation constant. We observe that the hot-electron high-energy plateau in the spectral rate becomes very weak for large momentum relaxation rates, according to the dependence $1/\gamma_p$. This tells us that metals with longer electronic mean free paths are preferable for efficient generation of high-energy electrons. In Figures 4-6, we show the data for gold and silver. We observe that, because of a longer mean free path, Ag NCs generate hot electrons much more efficiently than gold ones. The reason is that, in Ag NCs, the numbers of electrons with low excitation energies in a plasmonic wave is much smaller. The ratio for the efficiency of hot electron generation for the two considered metals is $\gamma_{p,Au}/\gamma_{p,Ag} \sim 4$.

In the next step, it is worthwhile to perform an analysis with the dissipation rates associated with low- and high-energy carriers in a plasmonic wave in Au and Ag NCs. Figure 6 shows the analysis for the rates

$$Rate_{low-energy}, \quad Rate_{high-energy}$$
$$Q_{low-energy}, \quad Q_{high-energy}$$

which are given by eqs 31 and 32. We see the following picture: The effect of the size on the ratio $Rate_{high-energy}/Rate_{low-energy}$ is dramatic. For photochemistry applications, we are, of course, interested in producing a larger number of high-energy electrons that can be used for reactions. Already for small NCs, the low-energy electrons start to dominate the carrier generation rates. In particular:

$$\frac{Rate_{high-energy}}{Rate_{low-energy}} < 0.3 \text{ for } a_0 > 5nm \text{ (Au-sphere)}$$
$$\frac{Rate_{high-energy}}{Rate_{low-energy}} < 0.3 \text{ for } a_0 > 15nm \text{ (Ag-sphere)} \quad (47)$$

The above ratios strongly depend on the material system. In silver, the mean free path is longer than that in gold and the rate of generation of of high-energy electrons is larger (Figures 6a,b and eqs 47). Essentially, we observe that the ratios (eqs 47) are inversely proportional to the Drude (momentum) scattering rate, i.e. $ratio \propto 1/\gamma_p$.

The situation looks better, however, for the energy efficiencies. The reason is that a small number of high-energy electrons can contain more energy than a large number of Drude electrons with energies close to the Fermi level. As we mentioned in the first section of this paper, the total absorption in a NC can be split into three terms (eq 13)

$$Q_{tot} = Q_{classical} + Q_{hot-electrons}$$
$$Q_{classical} = Q_{Drude} + Q_{inter-band}$$

where $Q_{classical}$ should be calculated according to classical electrodynamics (eqs 45 above), whereas $Q_{hot-electrons}$ is the quantum term (eq 35 above). In the next step, we introduce a quantum parameter for the plasmon, which plays the role of an efficiency parameter,

$$QP_{plamson} = \frac{Q_{hot-electrons}}{Q_{tot}} \quad (48)$$

In Figure 6d, we see that the efficiency persists for essentially larger NC sizes:

$$QP_{plasmon} < 0.3 \text{ for } a_0 > 9nm \quad (\text{Au-sphere})$$
$$QP_{plasmon} < 0.3 \text{ for } a_0 > 40nm \quad (\text{Ag-sphere})$$

This property, that the power efficiency persists for larger NCs as compared to the ratio of the rates, can be understood analytically:

$$\frac{Rate_{high-energy}}{Rate_{low-energy,Drude}} \propto \frac{1}{a_0^2}$$

$$QP_{plamson} = \frac{Q_{hot-electrons}}{Q_{tot}} \propto \frac{1}{a_0}$$

Another parameter, that can be useful to evaluate the over-barrier injection efficiency of the NC (Figure 1b), is a quantum yield of generation of electrons with over-barrier energies,

$$QY_{plasmon} = \frac{Rate_{hot-electrons,\varepsilon > \Delta E_{bar}}}{Rate_{tot, photon\ absorption}} \approx QP_{plasmon} \frac{\hbar\omega - \Delta E_b}{\hbar\omega}.$$

The above formula gives a reliable estimate for this parameter, while a more precise result can be found numerically using the integral $Rate_{hot-electrons,\varepsilon > \Delta E_{barrier}} = \int_{\Delta E_{bar}}^{\infty} R_e(\varepsilon)d\varepsilon$. Figure 6d shows the calculated quantum yields for Au and Ag NCs using a typical Schottky barrier height, $\Delta E_b = 1eV$.

The absorption and hot-electron spectra of spherical NCs of a fixed size are plotted in Figures 6e,f. Here one can again see the quantum hot-electron contributions in the absorption spectra. Importantly, as it was pointed out in refs 28 and 40, the absorption spectra and the rates will always have somewhat different shapes. For example, in the blue spectral interval for both Au and Ag NCs, the intraband hot-electron generation rate does not show the typical features of the interband absorption processes. For the Au case, we always observe in our calculations a small red shift of the generation rate as compared to the absorption spectrum (Figure 6e). A physical explanation for this comes from considering the interband term in the absorption. When we add the interband term to the intraband contribution, the plasmonic peak in the absorption spectrum gets slightly blue-shifted since the interband term is a strongly increasing function of photon frequency in the blue and UV intervals. For the Ag NCs, this effect is weaker and the plasmonic peaks for the absorption and generation rate are almost identical.

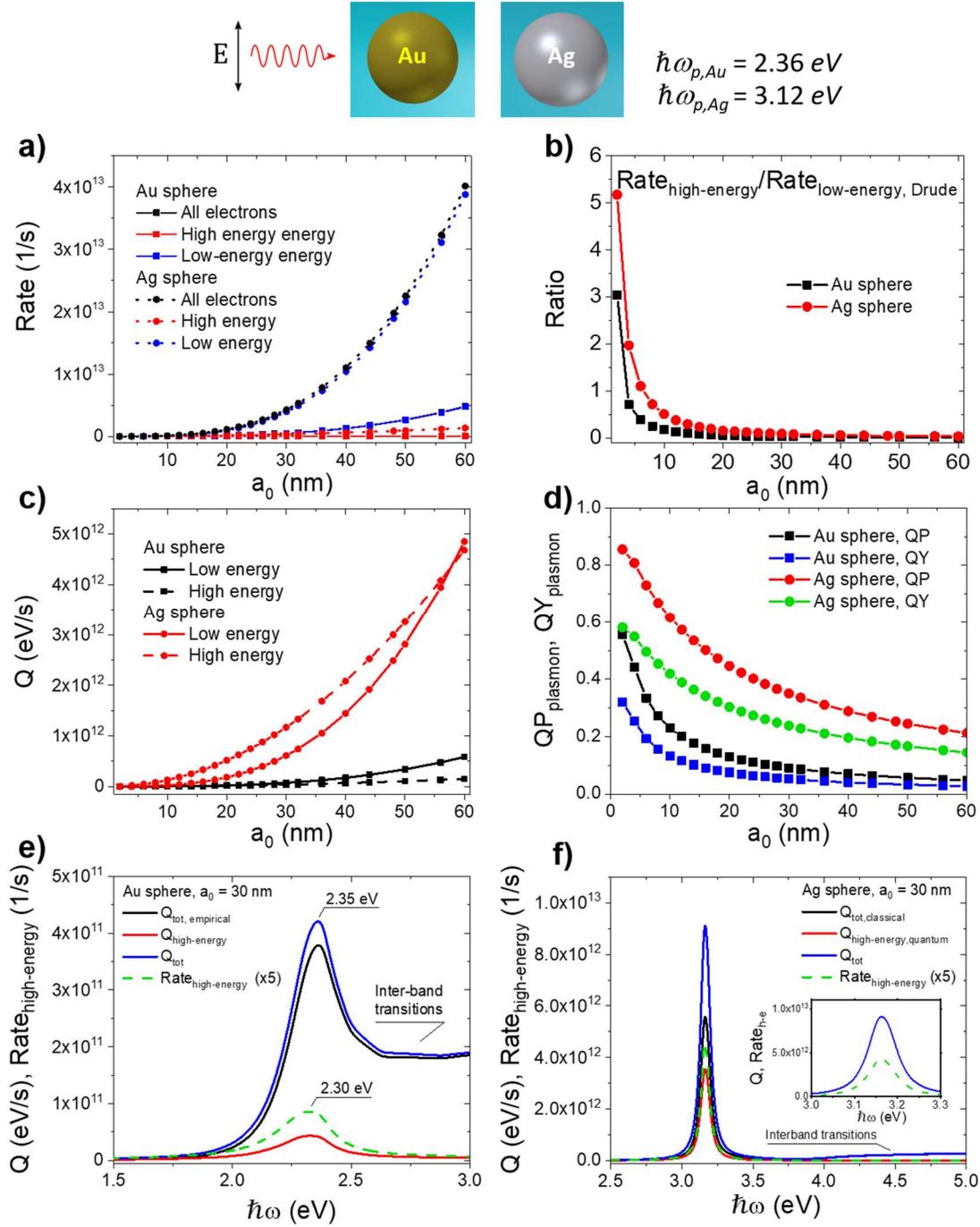

**Figure 6. (a)** Comparison between the rates of generation of low- and high-energy electrons in the Au and Ag NCs. **(b)** Ratios between the rates of generation of high- and low-energy electrons for the Au and Ag material systems. **(c)** Dissipations related to the generations of low- and high-energy electrons. **(d)** Efficiencies and quantum yields of hot-electron generation for the Au and Ag NCs. The quantum parameter (QP) gives the total efficiency of generation of high-energy electrons, whereas the quantum yield (QY) provides us with an efficiency of over-barrier generation; the barrier height in our calculation is 1 eV. **(e,f)** Absorptions in the units of eV/s and generation rates for 30nm Au and Ag spheres. For the Au case, one can see a shift between the total absorption and the rate. Also, the absorption and the rate behave differently in the blue spectral interval because of the interband transitions in these noble metals. The incident flux is $I_0 = 3.6 \cdot 10^3 W/cm^2$.

**Analytical equations for the hot-electron rates in spherical NCs.** For large spherical NCs, we can easily derive analytical equations for the energy-distributions and rates. These analytical equations are really convenient since, for large NC sizes, the number of involved single-particle transitions is big, and any exact calculations would require impracticably long computing times. The spectrum for the rate of generation of hot electrons has the following analytical form:

$$R_{e,high-energy,sphere}(\varepsilon) = Sign[\varepsilon - E_F] \cdot \frac{2}{\pi^2} \times \frac{e^2 E_F^2}{\hbar} \frac{1}{(\hbar\omega)^4} \frac{\pi}{3} a_0^2 \left| \frac{3\varepsilon_0}{2\varepsilon_0 + \varepsilon_{metal}} \right|^2 \frac{2\pi}{c_0 \sqrt{\varepsilon_0}} I_0, \quad \hbar\omega > |\varepsilon - E_F| > \delta E_{crit}.$$

(49)

It was derived in ref 46 and we repeat the derivation in Supporting Information of this article. Figures 4 and 5 show that the analytical expression in eq 49 works very well for spherical NCs. Then, the total rate of hot-electron production in a NC can be obtained in the following form:[46]

$$Rate_{high-energy,total} = \frac{2}{\pi^2} \times \frac{e^2 E_F^2}{\hbar} \frac{\hbar\omega - \delta E_{crit}}{(\hbar\omega)^4} \frac{\pi}{3} a_0^2 \left| \frac{3\varepsilon_0}{2\varepsilon_0 + \varepsilon_{metal}} \right|^2 E_0^2. \quad (50)$$

The above equation is also derived in Supporting Information as eq. S11. Correspondingly, the non-equilibrium population of electrons in the plasmonic wave within the first, simplified, approach (see prior discussion) reads

$$\delta n_{high-energy,sphere}(\varepsilon) = \tau_\varepsilon \cdot R_{e,high-energy,sphere}(\varepsilon) =$$
$$= Sign[\varepsilon - E_F] \frac{2}{\pi^2} \times \frac{e^2 E_F^2}{\gamma_\varepsilon} \frac{1}{(\hbar\omega)^4} \frac{\pi}{3} a_0^2 \left| \frac{3\varepsilon_0}{2\varepsilon_0 + \varepsilon_{metal}} \right|^2 \frac{2\pi}{c_0 \sqrt{\varepsilon_0}} I_0, \quad \hbar\omega > |\varepsilon - E_F| > \delta E_{crit}.$$

The quantum energy dissipation can be found by integrating the rate (see eq S12 in Supporting Information):

$$Q_{hot-electrons} = \int_{|\varepsilon - E_F| > \delta E_{crit}} \varepsilon \cdot R_{e,high-energy,sphere}(\varepsilon) \cdot d\varepsilon = \frac{2}{\pi^2} \times \frac{e^2 E_F^2}{\hbar} \frac{(\hbar\omega)^2 - \delta E_{crit}^2}{(\hbar\omega)^4} \frac{\pi}{3} a_0^2 \left| \frac{3\varepsilon_0}{2\varepsilon_0 + \varepsilon_{metal}} \right|^2 E_0^2$$

$$(51)$$

Figure S3a shows the characterization of the critical energy $\delta E_{crit}$, where the Drude electron part of the distribution turns into the quantum hot electrons. A good approximation for this energy reads:

$$\delta E_{crit} = \sqrt{(b_1 \frac{\hbar v_F \pi}{a_0})^2 + (b_2 k_B T_0)^2}, \quad (52)$$

where $T_0$ is the temperature of quasi-equilibrium of the electron gas, taken as the room temperature. We see that, for small NCs, this energy is fundamentally quantum in nature and, for large NCs, this parameter is given by the broadening of the Fermi distribution.

The low-energy (Drude) part of the spectral functions for large NCs (when $b_2 k_B T_0 > \hbar v_F \pi / a_0$) has the classical form:

$$R_{e,low-energy,sphere}(\varepsilon) = Q_{Drude,intra-band} \cdot \frac{d^2 f_F}{d\varepsilon^2},$$

$$Q_{Drude,intraband} = \frac{\gamma_p \omega_p^2}{\omega^2} \frac{1}{2\pi} V_{NC} \left| \frac{3\varepsilon_0}{2\varepsilon_0 + \varepsilon_{metal}} \right|^2 E_0^2 \qquad (53)$$

We note that the Drude dissipation rate in the above expression (53) was introduced earlier, in eq 45. Figure S3b shows again excellent agreement between the spectrum from a time-expensive computation and a simple analytical result (53). Another important piece of information is shown in Figure S3. The dissipation due to the low-energy electrons in large NCs should be, of course, equal to the classical Drude heating rate:

$$\int_{-\delta E_{crit}}^{\delta E_{crit}} \varepsilon \cdot R_{e,low-energy,sphere}(\varepsilon) \cdot d\varepsilon = Q_{Drude,intra-band} = \frac{\gamma_p \omega_p^2}{\omega^2} \frac{1}{2\pi} V_{NC} \left| \frac{3\varepsilon_0}{2\varepsilon_0 + \varepsilon_{metal}} \right|^2 E_0^2. \qquad (54)$$

In Figure S3, we can see a good overall agreement between the numerical integral over energy and the classical expression for Drude heating (left-hand and right-hand sides in eq 54). Some discrepancy in Figure S3 comes from the fact the value for the critical energy $\delta E_{crit}$ has some uncertainty.

Using the above analytical approaches, one can see that the ratio is directly proportional to the mean free path of an electron, $l_{mfp}$, in a particular metal:

$$\frac{Rate_{high-energy}}{Rate_{low-energy,Drude}} = const \cdot \frac{l_{mfp}}{a_0} \frac{v_F/a_0}{\omega}, \quad (55)$$

The Supporting Information gives the derivation for the estimate shown in eq 55. Regarding the energy efficiencies, the ratio for large NCs is given by

$$\frac{Q_{high-energy}}{Q_{low-energy,Drude}} = \frac{\gamma_{hot-electrons}}{\gamma_p} = const' \cdot \frac{l_{mfp}}{a_0} \quad (56)$$

The different dependence on size for eqs 55 and 56 is very interesting. Both equations have the physically-important parameter describing the ballistic motion of an electron, $l_{mfp}/a_0$. Of course, when $l_{mfp} \geq a_0$ an electron in a NC "feels" the walls and the quantum corrections due to the scattering by the surfaces should appear. However, this argument is not sufficient to explain the size-dependence of the ratio $Rate_{high-energy}/Rate_{low-energy,Drude}$ (eq 55). The additional and important parameter $\frac{v_F/a_0}{\omega}$ in the equation for the ratio of the rates appears because the matrix elements that contribute to the generation of hot electrons (eq 29) in most NCs are small. This is so because the single-particle wave functions oscillate with high spatial frequency and, for large energy differences between $\varepsilon$ and $\varepsilon'$, the integrals in eq 29 result in rather small values. Another way to look at this small parameter is to say that the effect of non-conservation of linear momentum in relatively large NCs is weak. Basically, the effects that we collect in eqs 55 and 56 come from breaking the Bloch theorem derived for an ideal periodic crystal.

**The importance of the material system: Au vs Ag.** We now comment on the striking difference between the excitation rates for Au and Ag NCs in Figure 6. Overall, a sharper plasmon resonance will produce more hot electrons with high energies,[46] since the quantum effects become enhanced in NCs with a long mean free path. We indeed see that for both ratios:

$$\frac{Rate_{high-energy}}{Rate_{low-energy,Drude}} \propto l_{mfp}, \quad \frac{Q_{high-energy}}{Q_{low-energy,Drude}} \propto l_{mfp}.$$

This is not simply caused by a stronger field enhancement and a stronger quantum absorption inside Ag NCs as compared to the Au case. In the ratios $Rate_{high-energy} / Rate_{low-energy,Drude}$ and $Q_{high-energy} / Q_{low-energy}$, both denominators and numerators are enhanced. However, these ratios are typically larger in Ag NCs than those in the Au system because of the appearance of stronger quantum effects in metals with a longer mean free path of electron. In Figures 6e,f, we again see that the quantum contribution for the Ag material system in the whole visible-light spectral interval is stronger than in the case of gold.

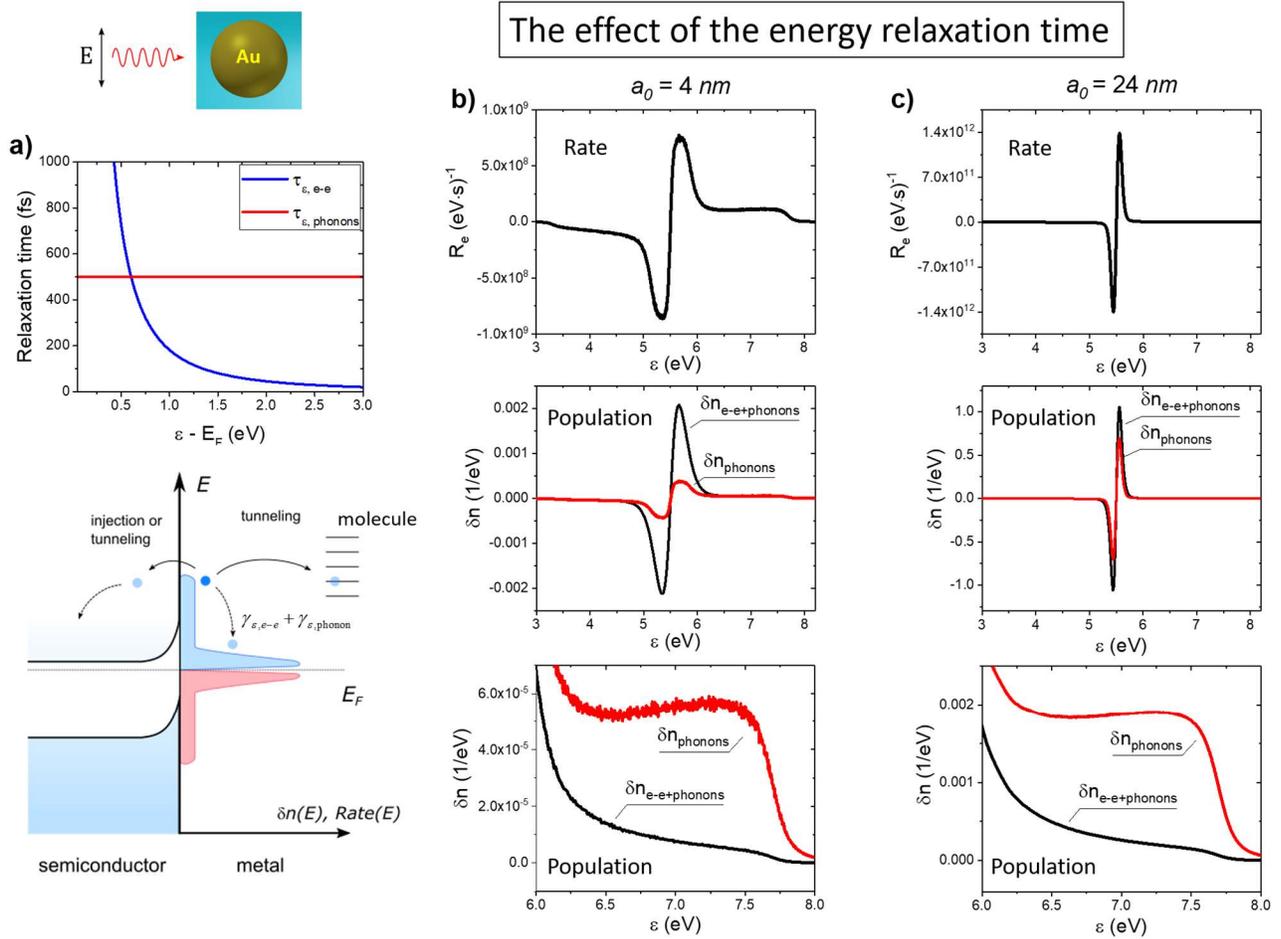

**Figure 7. (a)** The relaxation rates of single electrons due to phonon and electron-electron scattering mechanisms, used in the calculations. **(b,c)** Generation rates and populations of excited electrons for Au NCs with two sizes (4nm and 24nm). The steady-state populations are calculated within the two approaches described in the text. One approach involves only phonon-assisted scattering and the other includes both phonon-assisted and electron-electron scattering channels. The incident flux is $I_0 = 3.6 \cdot 10^3 W/cm^2$.

**The role of the energy relaxation mechanisms.** Figure 7 shows a collection of results for the steady-state distribution of plasmonic electrons in a small Au NP (4nm) and in a larger one (24nm). Panel a of this figure shows the relaxation times in femtoseconds, accompanied by a schematic representation of the relaxation mechanisms, whereas panels b,c present numerical results for the rates and the electronic population calculated using two semi-analytical approaches (as discussed earlier in this paper). The effects of the hot-electron relaxation mechanism are particularly relevant because the characteristic e-e collision time is very fast for high-energy electrons and absolutely dominates the relaxation of energetic carriers (Figure 7a). Therefore, this mechanism efficiently "removes" carriers from high energies and then "place" these carriers at small energies. We see this striking behavior in panels b,c in Figure 7.

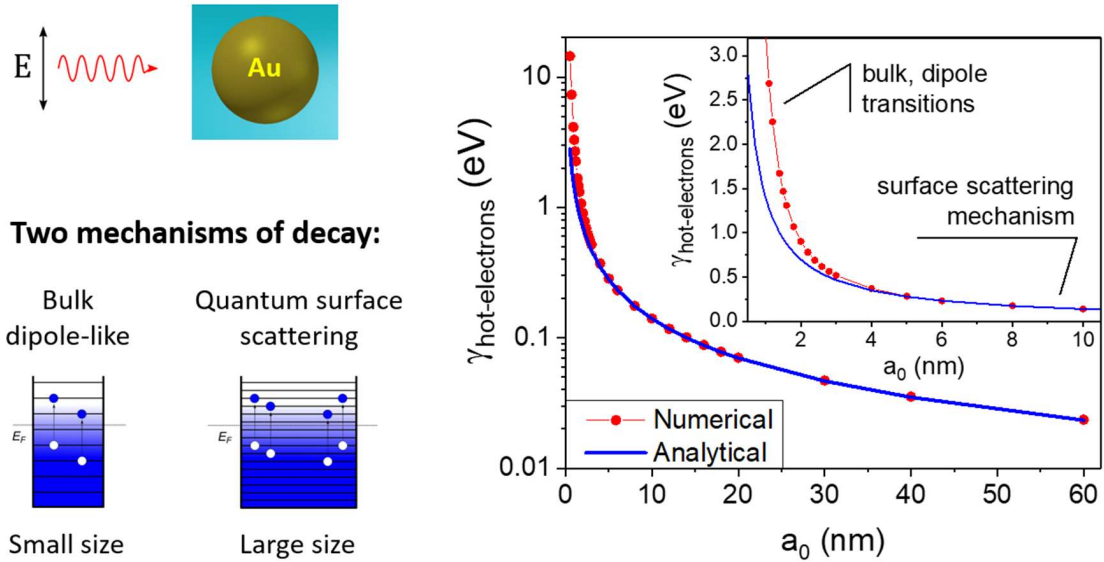

**Figure 8.** Rate of plasmon decay due to collisions with the surface, shown in semi-logarithmic scale and data for small NCs in linear scale in the inset; $\hbar\omega = 2.36\ eV$. Diagrams: Model of an Au NC and two mechanisms of plasmon decay.

**Plasmonic lifetimes coming from the hot electron generation in plasmonic spheres.** At the beginning of this paper, we discussed the origin of the relaxation rates for plasmons. Each type of absorption creates a relaxation mechanism for a plasmon. The phenomenological equation of energy conservation for the plasmon under the CW illumination tells us that the plasmonic decay rate due to hot-electron generation has the form:

$$\gamma_{hot-electrons} = \frac{Q_{hot-electrons}}{E_{plasmons}}.$$

However, it is more practical to use a different equation to express this rate. We now neglect the interband transitions and consider the Drude metal, for simplicity. In the next step, we note that the Drude relaxation rate should be given by

$$\gamma_{Drude} = \gamma_p = \frac{Q_{Drude,intra-band}}{E_{plasmons}}.$$

If such, the ratio between them becomes:

$$\frac{\gamma_{hot-electrons}}{\gamma_{Drude}} = \frac{Q_{hot-electrons}}{Q_{Drude,intra-band}}$$

This expression does not contain the stored plasmonic energy, $E_{plasmons}$. Instead, this equation includes the typically computed dissipation $Q_{Drude,intra-band}$. Then, we obtain a convenient equation for the plasmonic decay rate for the hot-electron path (Figure 2a)

$$\gamma_{hot-electrons} = \gamma_{Drude} \frac{Q_{hot-electrons}}{Q_{Drude,intra-band}}.$$

For large spherical NCs, we immediately arrive to the following analytical equation:

$$\gamma_{hot-electrons} = \frac{3}{2} \frac{\hbar v_F}{a_0}.$$

To get this result we used the analytical eqs 51 and 53 and assuming $\hbar\omega \gg \delta E_{crit}$. This result is in agreement with a number of previous calculations for the surface-induced rates of plasmonic decay.[54,56–58] This simple analytical limit for large sizes is well reproduced in our numerical calculations for Au presented in Figure 8. Analogous calculations for Ag are given in the Supporting Information (Figure S5). Again, the size dependences in Figures 8 and S5 were averaged over a reasonable NC size range, to obtain smooth curves. However, our calculations provide another interesting result: the plasmonic rates for small sizes (~ 1.5nm) in Figures 8 and S5 start to deviate significantly from the well-known Kreibig's dependence, $\propto a_0^{-1}$. By analyzing the quantum transitions in small NCs, we reached the conclusion that this deviation comes from very strong dipolar-like transitions in small NCs with sizes ~ 1-2nm. For such small sizes, we already satisfy the following condition: $\hbar v_F (\pi / a_0) \sim \hbar \omega$. This means that the major dipole-like transitions in a NC ($\Delta n_r \sim 1, \Delta l = \pm 1, \Delta m = 0, \pm 1$) become resonant for the process of plasmon decay, where $n_r$ is the radial quantum number of an electron in a spherical well. This numerical observation can be important for understanding the behavior of the plasmon peak in ultra-small NCs. It was observed in such ultra-small NCs (1-2nm) that the plasmon peak vanishes when a NC size becomes smaller than some critical value.

**Analytical equations for the efficiencies.** We can also derive analytical equations for the efficiency of hot-electron generation. This parameter, which is given by eq 48 can be also expressed via the plasmon decay rates:

$$QP_{plasmon} = \frac{\gamma_{hot-electrons}}{\gamma_{Drude} + \gamma_{inter-band} + \gamma_{hot-electrons}}.$$

For a spherical NC, we can use the convenient results for the plasmonic rates, obtained in a previous section: $\gamma_{hot-electrons} / \gamma_{Drude} = (3/2) \cdot (v_F / a_0) / \gamma_p$. Then, we obtain:

$$QP_{plamson} = \frac{1}{1 + \frac{2}{3}\frac{a_0}{l_{eff}}},$$

where $\gamma_{inter-band} = \text{Im}(\varepsilon_{inter-band})\hbar\frac{\omega^3}{\omega_p^2}$ and $l_{eff}^{-1} = (\gamma_p + \gamma_{inter-band})/(\hbar v_F)$. To obtain the Drude model results, we should remove the inter-band broadening and the results will be expressed via the mean free path: $l_{eff} \to l_{mfp} = v_F \cdot \tau_p$. The efficiency decreases with the increasing size of a NC, since the hot-electron effect is a surface effect. For large NCs, the efficiency behaves as

$$QP_{plamson} \sim \frac{3}{2}\frac{l_{eff}}{a_0}.$$

**Analytical results for NCs with arbitrary shapes.** The approach of integrating the enhanced normal electric field over the surface can be applied to NCs with an arbitrary shape.[46] Such NCs can be slabs, spheres, nanorods, or others. Since hot electrons are produced in nearly flat distributions, we can write down:

$$Q_{hot-electrons} = \frac{\left((\hbar\omega)^2 - \delta E_{crit}^2\right)}{(\hbar\omega - \delta E_{crit})} \cdot Rate_{high-energy} \approx \hbar\omega \cdot Rate_{high-energy},$$

Then, using integration over the NC surface, we obtain:

$$Rate_{high-energy} \approx \frac{2}{\pi^2} \times \frac{e^2 E_F^2}{\hbar} \frac{1}{(\hbar\omega)^3} \int_{S_{NC}} |E_{normal}(\theta,\varphi)|^2 \, ds$$

$$Q_{high-energy} \approx \frac{2}{\pi^2} \times \frac{e^2 E_F^2}{\hbar} \frac{1}{(\hbar\omega)^2} \int_{S_{NC}} |E_{normal}(\theta,\varphi)|^2 \, ds \qquad (57)$$

The derivations for these equations are simple and can be found both in ref 46 and in the Supporting Information. It is shown below that this approach is not applicable for NCs with very sharp tips, for which the surface integration does not give a physically relevant result. In NCs with hot spots, the locality of generation of hot carrier on the surface becomes broken because the system acquires a small characteristic size. This small size can be the small radius of curvature characterizing the edges and vertices of a nanocube.

We should note that a similar approach of surface integration was previously used in ref [69] to calculate the total quantum dissipation in a NC arising from surface scattering of electrons. The dissipation rate per unit surface was calculated in ref [69] using time-dependent perturbation theory.[70,71] In our quantum kinetic theory, we apply the surface integration method to calculate the energy distributions of hot electrons for NCs with smooth surfaces. Through this method we can, for example, answer the key question how many hot electrons are excited in a NC above the barrier energy. Moreover, our linear-response theory, which is based on the density-matrix master-equation formalism with two or more relaxation times, allows us to see and calculate the whole spectrum of excited electrons for low and high excitation energies. In the electron distribution function, we recognize both Drude electrons and energetic electrons generated by the surface-assisted quantum transitions.

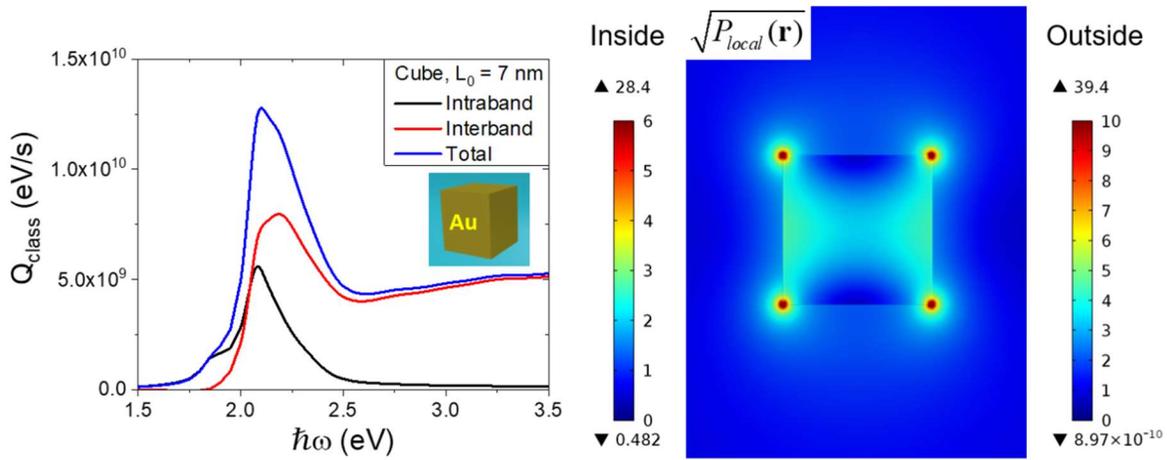

**Figure 9. (a)** Spectrum of classical dissipation of light energy in an Au nanocube with sharp edges. **(b)** Color map for the electric field enhancement of an optically-driven nanocube; this map has different color scales for the internal and external fields. One can see strong hot spots at the vertices of the nanocrystal.

## The case of hot spots: nanocubes and dimers.

**General remarks and comparison with the nanosphere geometry.** In nanocrystals with a smooth surface, the generation rate in the plateau regions of the function $R_{e,high-energy,sphere}(\varepsilon)$ should be well approximated by the surface integral:

$$R_{e,high-energy,sphere}(\hbar\omega > |\varepsilon - E_F| > \delta E_{crit}) \approx Sign[\varepsilon - E_F] \cdot f_1(\omega) \int_{S_{NC}} |E_{normal}(\theta,\varphi)|^2 \, ds \quad (58)$$

From the above equation (eq 58), it is clear that the rate is proportional to the surface field-enhancement factor $P_{enh,s}$. Then, the total rate of hot-electron generation and the quantum dissipation should also be proportional to the surface integrals:

$$Rate_{high-energy} = f_2(\omega) \int_{S_{NC}} |E_{normal}(\theta,\varphi)|^2 \, ds \propto S_{NC} \cdot P_{enh,s}$$

$$Q_{high-energy} = f_3(\omega) \int_{S_{NC}} |E_{normal}(\theta,\varphi)|^2 \, ds \propto S_{NC} \cdot P_{enh,s} \quad (59)$$

In the above equations, the functions $f_{1,2,3}(\omega)$ bundle the extra factors and dependency on $\omega$ shown in eqs 49 to 51. One can see that all hot-electron properties of a NC with smooth surfaces depend on the surface integral of the normal field. We see this property for a single nanosphere and for dimer (see the results below). However, in the case of nanocubes we did not find such simple dependence on the surface integral, since the generation of hot carriers and the quantum hot-electron dissipation are not purely local surface effects. The electric field inside a nanocube becomes strongly amplified and inhomogeneous in the hot-spot regions near the cube vertexes (Figure 9).

Computationally, quantum calculations for nanocubes and nanoparticle dimers are much more demanding as compared to the case of a sphere. The reasons lie in a large number of involved matrix elements and in the lifted selection rules for quantum transitions in these NCs. In particular, we have used $\sim 1.5 \cdot 10^8$ matrix elements for the largest computed nanocube ($L = 7nm$) and $\sim 2.9 \cdot 10^6$ elements for the largest computed dimer ($a_0 = 6nm$).

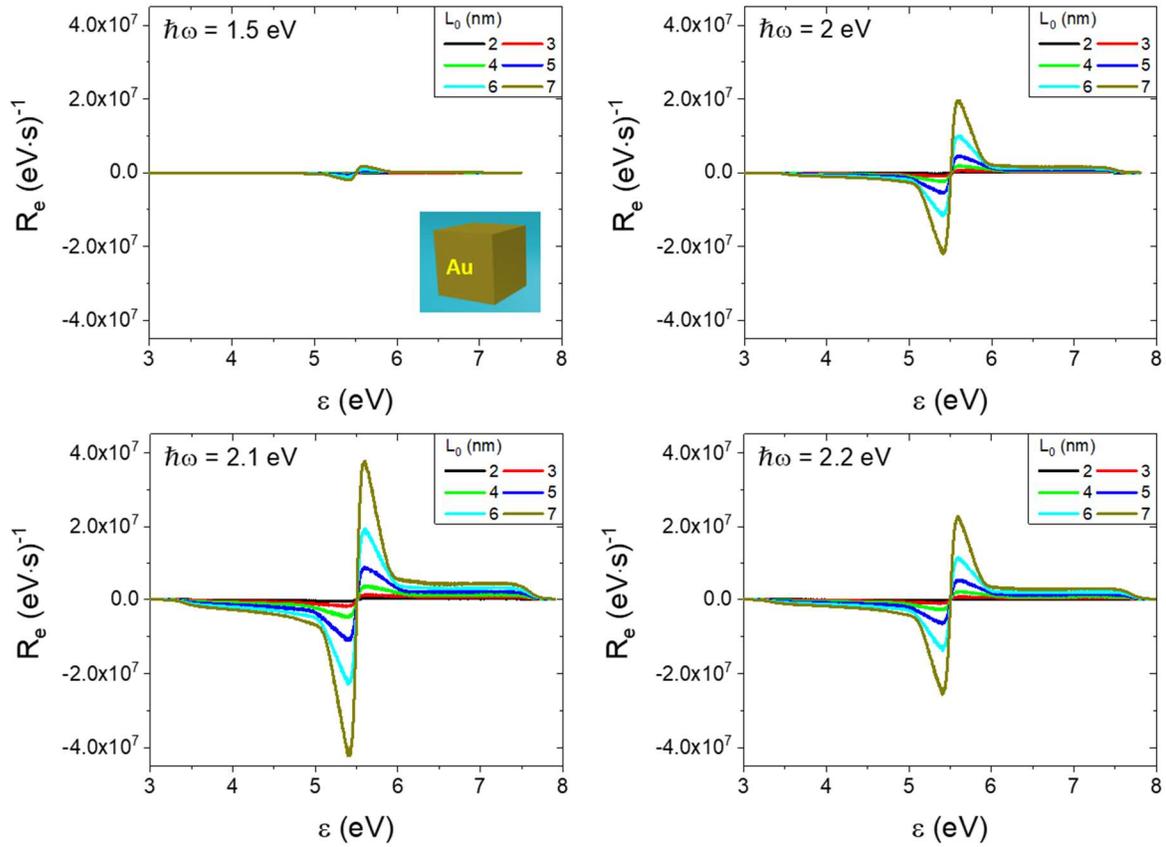

**Figure 10.** Rates of generation of excited electrons in Au nanocubes of different sizes for four different photon energies. All panels share the same scale, which makes easier to appreciate the resonant response of the system. The incident flux is $I_0 = 3.6 \cdot 10^3 W/cm^2$.

**Nanocubes.** To model the nanocube system, we take the simplest wave functions of a cubic quantum well:

$$\psi_\mathbf{n} = \sqrt{\frac{2^3}{L_0^2 L_z}} sin(k_{n_x} x) sin(k_{n_y} y) sin(k_{n_z} z)$$

$$\mathbf{k_n} = (k_{n_x}, k_{n_y}, k_{n_y}) = \left(\frac{\pi}{L} n_x, \frac{\pi}{L} n_y, \frac{\pi}{L} n_z\right)$$

$$\varepsilon_\mathbf{n} = \frac{\hbar^2 \pi^2 (n_x^2 + n_y^2 + n_z^2)}{2 m_0 L^2}$$

where $L$ is the cube size. The quantum state of an electron in a nanocube is described by the numbers $\mathbf{n} = (n_x, n_y, n_z)$, where $n_\alpha = 1, 2, 3, \ldots$. The electric field inside a nanocube was computed numerically using the COMSOL software (Figure 9).

Figure 10 shows detailed results for the electron excitation rates in nanocubes. We again observe the general features of hot and Drude electrons – the positive and negative peaks at low excitation energies and the plateaus for hot electrons and holes. Since the sizes of the nanocubes in Figure 10 are relatively small, the shapes of the low-energy peaks for the Drude electrons are triangular (like in the model of a slab[29,40]) and have the essentially quantum nature originating from the quantum dipolar transitions with $\Delta n_z = \pm 1$. The plateaus come from transitions with $\left|\Delta n_z\right| \gg 1$. It is interesting to understand the nature of the high-energy plateaus in Figure 10. In the nanosphere and dimer (see below), the generation of high-energy electrons comes from the surfaces and the rate at the plateau is proportional to the surface area and the surface field enhancement factor: $R_{e,plateau}(\varepsilon) \propto S_{NC} \cdot P_{enh,s}$. This is not the case for the nanocube, since the generation of hot electrons is governed by the strong hot spots at the vertices.

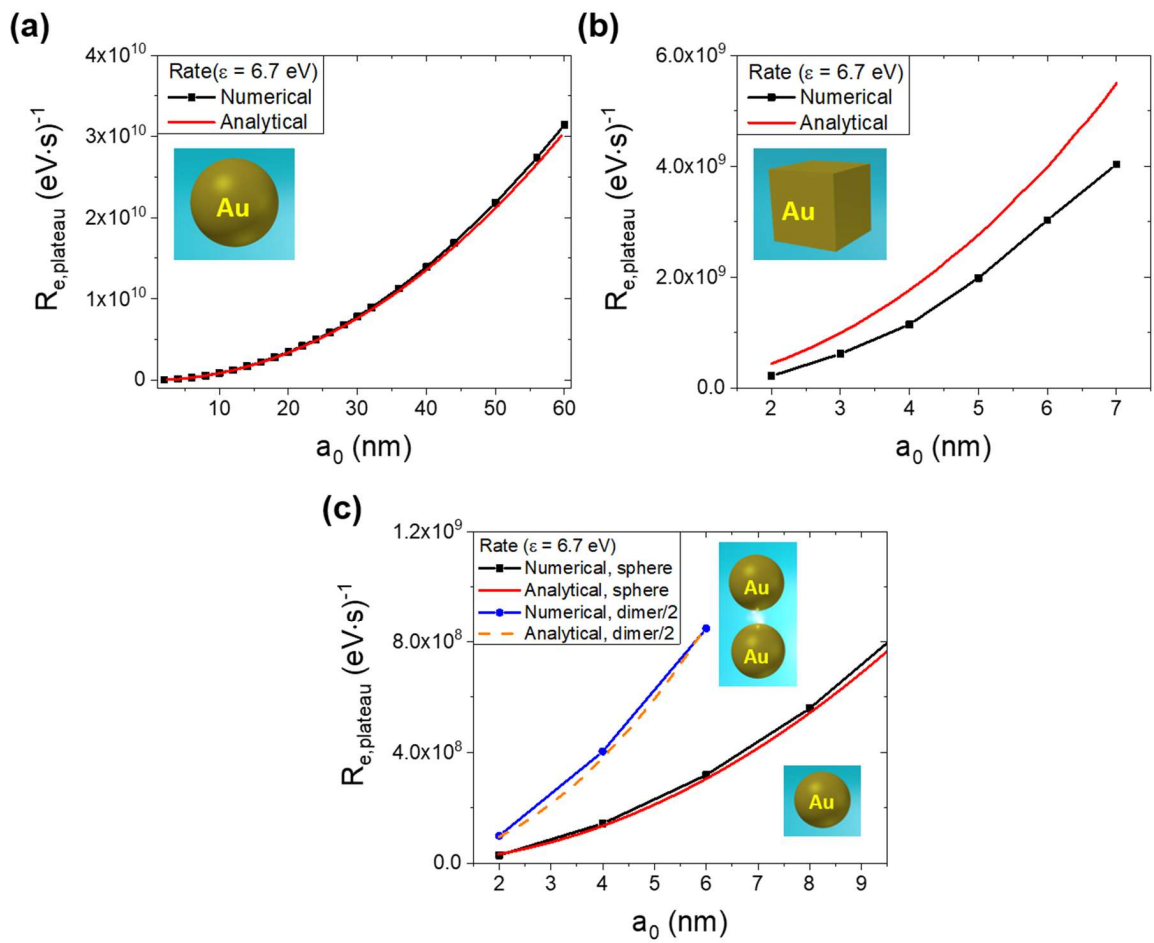

**Figure 11.** Rates of electron generation at the plateau ($\varepsilon = 6.7\, eV$) for three NC systems. The excitation energies are taken close to the plasmonic resonances of these systems. The graph shows both numerical data and analytical results coming from the surface integration (eq 58). We see very good agreement between the numerical and semi-analytical methods for the sphere and for the dimer, but not for the cube.

For the cube, the local analytical approach based on the surface integral fails (see Figure 11b). In Figure 11b we see that the semi-analytical equation for the rate based on the surface integral gives significantly larger values, whereas the surface integration approach works perfectly for the spheres and the dimers with hot spots (Figures 11a,c). The reason is that the local surface generation approach is not valid if the radius of curvature of a NC's surface is less than the characteristic sizes of the Fermi gas, such as the Fermi length and the Thomas-Fermi screening length. The Fermi length in Au and Ag is $k_F^{-1} \sim 0.8\,\text{Å}$. Thomas-Fermi screening length is calculated as

$$l_{TF} = \sqrt{\frac{6\pi e^2 n_0}{\varepsilon_{b,Drude} E_F}},$$

where $n_0$ is the electron density $\varepsilon_b$ is the background screening constant of the matrix. For gold, $l_{TF} \sim 1.8\,\text{Å}$ using $\varepsilon_{b,Drude} = 9.07$ from Table 1. Then, we see that our model for the cube geometry has too sharp vertices to satisfy the local approximation for the surface rate. We see the implications of this choice of geometry in Figure 11b.

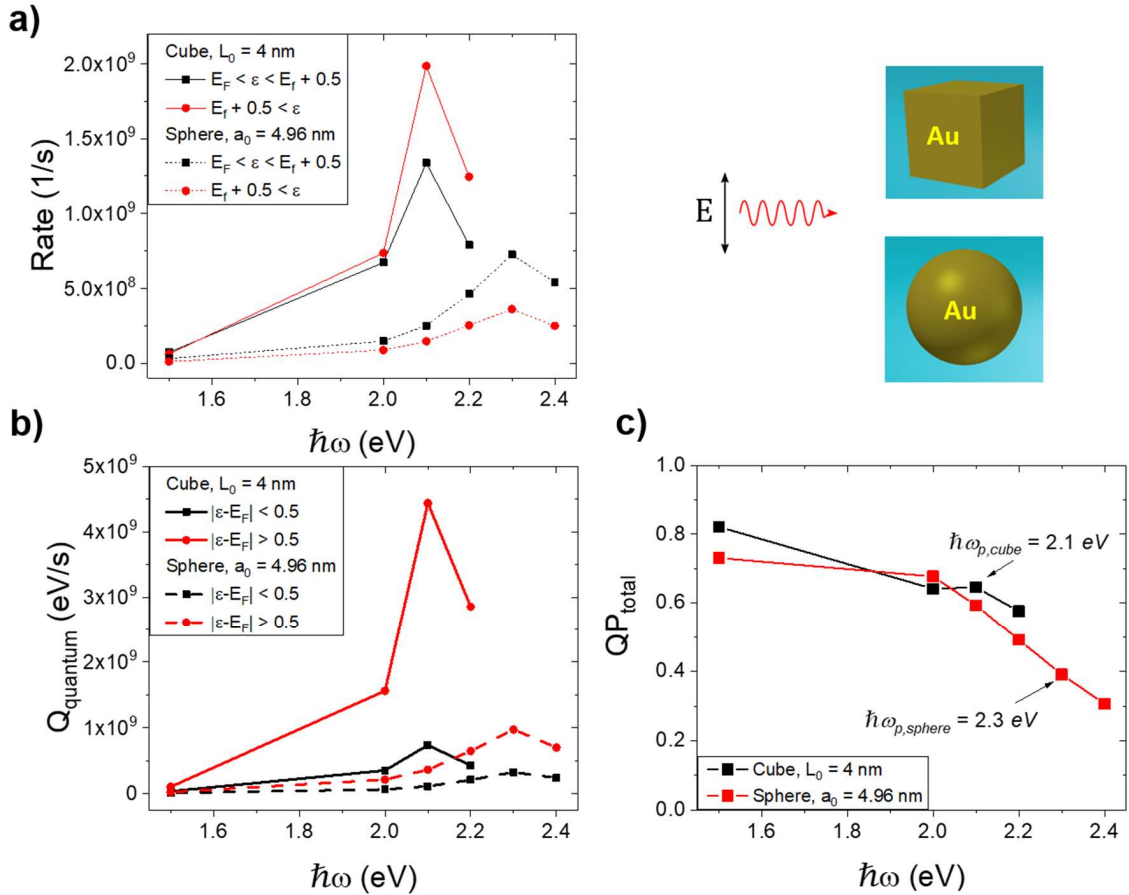

**Figure 12.** Properties of a nanocube as compared with a nanosphere. The sizes of the NCs were chosen so that their volumes are equal. **(a)** Total excitation rates for high-energy and low-energy electrons. **(b)** Quantum absorption related to the electron excitation for the nanocube and nanosphere. **(c)** Quantum parameter of plasmon, or efficiency of generation of hot electrons.

As expected, nanocubes can generate hot electrons more efficiently in comparison with nanospheres of the same volume, since they have stronger inner electromagnetic fields. This can be seen in Figure 12, where we show the spectra for the rates, both for a nanocube and a nanosphere with equal volumes. In addition, we can expect the quantum effect in nanocubes to be stronger since the electromagnetic fields are more strongly enhanced and are more inhomogeneous for this shape of NC. We indeed observe this property in Figure 12c: the quantum parameter of plasmon for the nanocube at its plasmon frequency is larger than that of the nanosphere.

**Plasmonic dimers.** To better understand the electronic responses in NCs with different shapes, we now look at the field enhancement factors (Figure 13). We see that the nanocube has clearly the larger surface fields of the three considered model systems. Then, we look at the rates of generation (Figure 14). The largest rate of generation is observed for the cubes, since they show the strongest field enhancement. We also observe that the nanosphere and the dimer are well described by the surface integrals and, in general, follow the mechanism of surface generation of hot carriers (Figures 11, 14b). Simultaneously, the response of the cube is strongly nonlocal and the rate does not follow the surface enhancement factor (Figure 14b). Basically, the analytical expression in eq 58 strongly overestimates the rate (Figure 14b). Therefore, we conclude that the cubes exhibit the mechanism of generation based on hot spots. In the dimer case, the hot spot also plays an important role and, consequently, the surface integral of the normal field is mostly determined by the area near the gap:

$$Rate_{high-energy} = Rate_{hot\ spot} + Rate_{rest} =$$

$$= f_2(\omega) \left( \int_{hot\ spot} |E_{normal}(\theta,\varphi)|^2 ds + \int_{rest} |E_{normal}(\theta,\varphi)|^2 ds \right).$$

The surface area of the hot spot in the dimer is relatively small (see Figure 13), but plays the leading role in the generation of hot electrons. We see this behaviors in Figures 13c,d. The surface integral for the normal electric field is now composed of two parts, $P_S = P_{S,hot\,spot} + P_{S,rest}$, and the hot spot region is taken as a polar area within the angle of 45°; the solid angle of the chosen hot-spot area is only the 4% of the total solid angle of the sphere, $4\pi$. At the plasmon resonance the hot-spot generates a similar number of hot electrons as the rest of the sphere (Figure 13d). Therefore, 50% of the generation rate is attributed to the hot spot, that is only 4% of the total surface area. Since the hot-spot area has an increased rate of generation of energetic electrons, such a spatial region will greatly contribute to the process of decay of plasmon, i.e. to the parameter $\gamma_{hot-electrons}$. This parameter can be also written as a sum $\gamma_{hot-electrons} = \gamma_{hot-spot} + \gamma_{rest}$, where the hot spot becomes the main source of decay of plasmon. We note that, for the case of a plasmonic cube, greatly increased numbers of energetic electrons due to the hot spots were found before in ref [40].

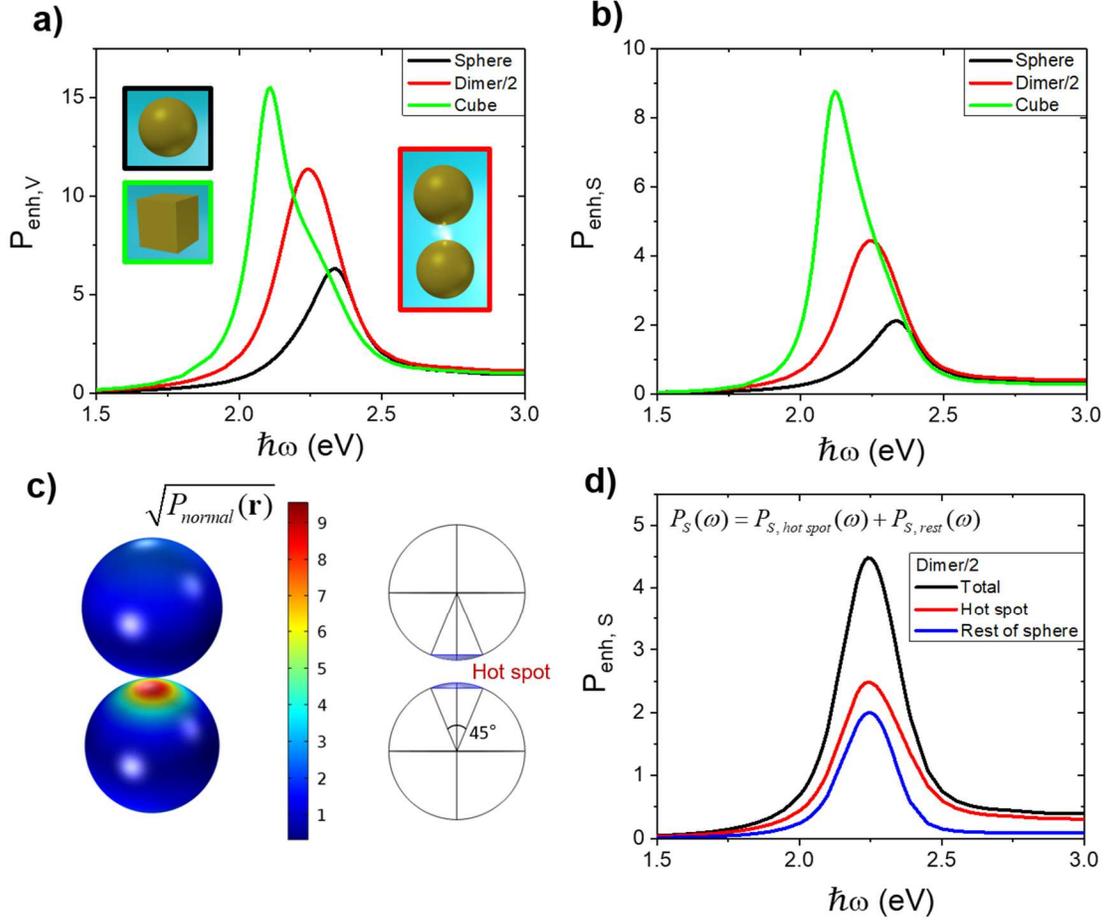

**Figure 13.** Field enhancement factors for three types of NCs used in the calculations. **(a)** Volume integral of the enhancement inside the NCs. **(b)** Surface integral of the inner enhancement at the NCs' surfaces. **(c)** Color map showing the function $\sqrt{P_{normal}(\mathbf{r})} = \sqrt{|E_{norm}(\theta,\varphi)|^2/E_0^2}$ at the dimer's surface, alongside a diagram showing the area of the hotspot. The calculated dimer has [the] a gap equal to $a_0/6$. **(d)** Dimer data from panel (b), separating the contribution of the hotspot and the rest of the sphere's surface.

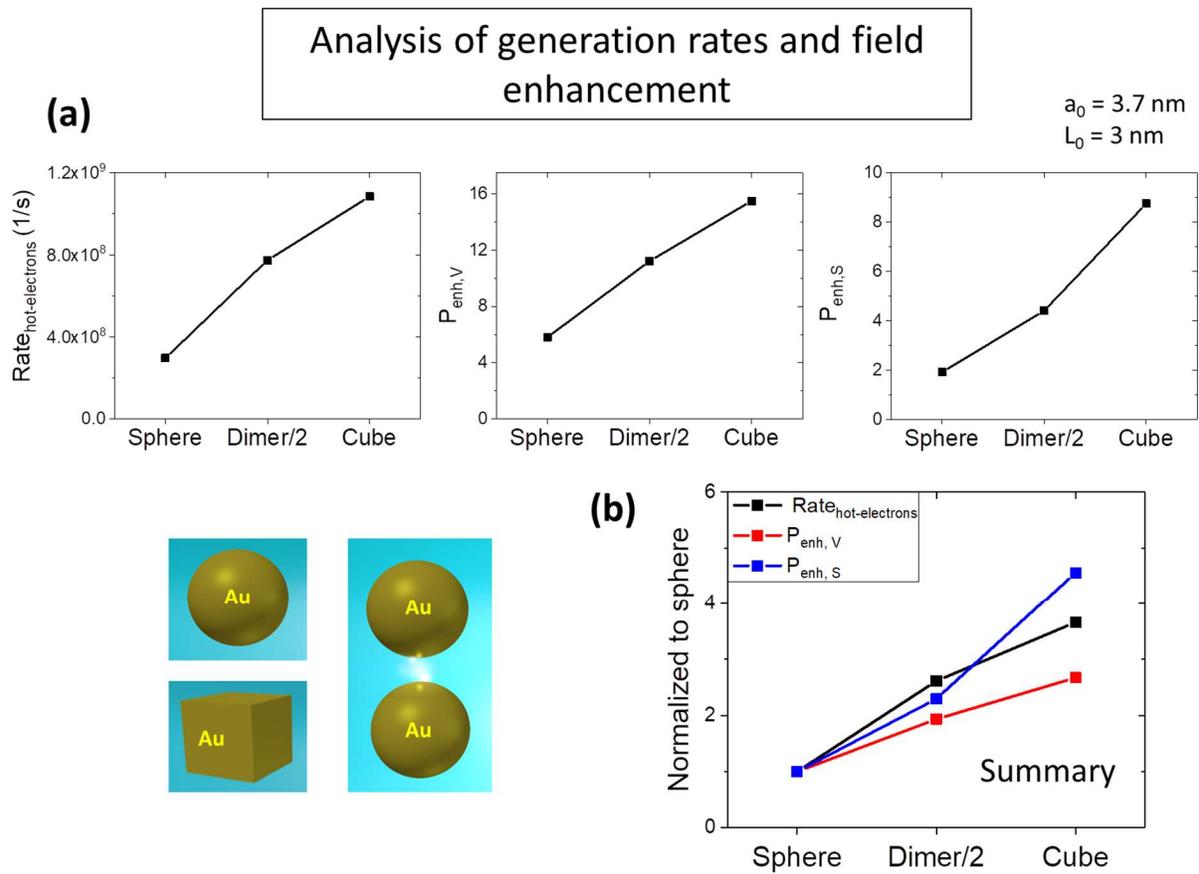

**Figure 14** **(a)** Rates of hot electron generation and enhancement factors for three types of NCs. The volume is kept the same for all NCs. For the dimer, we count the generation only in one sphere. **(b)** Results normalized to the values of the isolated sphere, showing the general trends.

# Summary and conclusion.

We have developed both numerical and analytical approaches for the problem of hot electron generation in NCs with and without hot spots. The plasmon decay in a metal NC can be viewed as a multi-channel process with two main channels: Classical (Drude) friction-like dissipation and generation of hot electrons. The latter is an interesting quantum process and originates from breaking conservation of linear momentum of an electron in a NC with surfaces and hot spots. For simple NC shapes, such as a sphere, we derived analytical equations for the rates and numbers of low-energy and high-energy electrons. These equations can be useful to obtain quick estimates for the numbers of hot electrons in realistic systems. The case of hot spots is very interesting for two reasons: (1) the hot-spot NCs generate hot carriers much more efficiently and (2) the generation may not be described by the surface integral. We identify three mechanisms of hot electron generation: (a) the normal surface generation mechanism that is typical for spheres; (b) the surface mechanism due to hot spots in nanoparticles with small gaps when the surface region with a hot spot governs the hot-electron generation; (c) Bulk hot spots near the sharp tips of a NC when the total rate does not follow the surface fields; this occurs in the cubes with sharp vertices. Optical generation of energetic electrons is an intrinsic property of any plasmonic system and, therefore, our results can be utilized for designing and understanding a variety of plasmonic and hybrid nanoscale devices.

**Supporting Information:**
Details regarding the calculations of electronic and optical properties of nanocrystals.

**ACKNOWLEDGEMENTS**
This work was supported by Volkswagen Foundation (Germany) and via the Chang Jiang (Yangtze River) Chair Professorship (China).

**Corresponding Authors**
*Contact details:** lvbesteiro@gmail.com, (740) 597-2661;  govorov@ohiou.edu, (740) 593-9430.

(16) Zhang, X.; Li, X.; Zhang, D.; Su, N. Q.; Yang, W.; Everitt, H. O.; Liu, J. Product Selectivity in Plasmonic Photocatalysis for Carbon Dioxide Hydrogenation. *Nat. Commun.* **2017**, *8*, 14542.
(17) Liu, L.; Zhang, X.; Yang, L.; Ren, L.; Wang, D.; Ye, J. Metal Nanoparticles Induced Photocatalysis. *Natl. Sci. Rev.* **2017**.
(18) Hou, B.; Shen, L.; Shi, H.; Kapadia, R.; Cronin, S. B. Hot Electron-Driven Photocatalytic Water Splitting. *Phys. Chem. Chem. Phys.* **2017**, *19*, 2877–2881.
(19) Wang, C.; Nie, X.-G.; Shi, Y.; Zhou, Y.; Xu, J.-J.; Xia, X.-H.; Chen, H.-Y. Direct Plasmon-Accelerated Electrochemical Reaction on Gold Nanoparticles. *ACS Nano* **2017**, *11*, 5897–5905.
(20) See, E. M.; Peck, C. L.; Guo, X.; Santos, W. L.; Robinson, H. D. Plasmon-Induced Photoreaction of O-Nitrobenzyl Based Ligands under 550 Nm Light. *J. Phys. Chem. C* **2017**, *121*, 13114–13124.
(21) Boriskina, S. V.; Ghasemi, H.; Chen, G. Plasmonic Materials for Energy: From Physics to Applications. *Mater. Today* **2013**, *16*, 375–386.
(22) Narang, P.; Sundararaman, R.; Atwater, H. A. Plasmonic Hot Carrier Dynamics in Solid-State and Chemical Systems for Energy Conversion. *Nanophotonics* **2016**, *5*, 96.
(23) Mangold, M. A.; Weiss, C.; Calame, M.; Holleitner, A. W. Surface Plasmon Enhanced Photoconductance of Gold Nanoparticle Arrays with Incorporated Alkane Linkers. *Appl. Phys. Lett.* **2009**, *94*, 161104.
(24) Chalabi, H.; Schoen, D.; Brongersma, M. L. Hot-Electron Photodetection with a Plasmonic Nanostripe Antenna. *Nano Lett.* **2014**, *14*, 1374–1380.
(25) Pelayo García de Arquer, F.; Mihi, A.; Konstantatos, G. Molecular Interfaces for Plasmonic Hot Electron Photovoltaics. *Nanoscale* **2015**, *7*, 2281–2288.
(26) Li, W.; Coppens, Z. J.; Besteiro, L. V.; Wang, W.; Govorov, A. O.; Valentine, J. Circularly Polarized Light Detection with Hot Electrons in Chiral Plasmonic Metamaterials. *Nat. Commun.* **2015**, *6*, 8379.
(27) Li, W.; Valentine, J. G. Harvesting the Loss: Surface Plasmon-Based Hot Electron Photodetection. *Nanophotonics* **2016**, *6*, 177.
(28) Govorov, A. O.; Zhang, H.; Gun'ko, Y. K. Theory of Photoinjection of Hot Plasmonic Carriers from Metal Nanostructures into Semiconductors and Surface Molecules. *J. Phys. Chem. C* **2013**, *117*, 16616–16631.
(29) Govorov, A. O.; Zhang, H. Kinetic Density Functional Theory for Plasmonic Nanostructures: Breaking of the Plasmon Peak in the Quantum Regime and Generation of Hot Electrons. *J. Phys. Chem. C* **2015**, *119*, 6181–6194.
(30) DeSario, P. A.; Pietron, J. J.; Dunkelberger, A.; Brintlinger, T. H.; Baturina, O.; Stroud, R. M.; Owrutsky, J. C.; Rolison, D. R. Plasmonic Aerogels as a Three-Dimensional Nanoscale Platform for Solar Fuel Photocatalysis. *Langmuir* **2017**.
(31) Manjavacas, A.; Liu, J. G.; Kulkarni, V.; Nordlander, P. Plasmon-Induced Hot Carriers in Metallic Nanoparticles. *ACS Nano* **2014**, *8*, 7630–7638.
(32) Zhang, Y.; Yam, C.; Schatz, G. C. Fundamental Limitations to Plasmonic Hot-Carrier Solar Cells. *J. Phys. Chem. Lett.* **2016**, *7*, 1852–1858.
(33) Sundararaman, R.; Narang, P.; Jermyn, A. S.; Goddard III, W. A.; Atwater, H. A. Theoretical Predictions for Hot-Carrier Generation from Surface Plasmon Decay. *Nat. Commun.* **2014**, *5*, 5788.
(34) Naik, G. V.; Dionne, J. A. Photon Upconversion with Hot Carriers in Plasmonic Systems. *Appl. Phys. Lett.* **2015**, *107*, 133902.
(35) White, T. P.; Catchpole, K. R. Plasmon-Enhanced Internal Photoemission for Photovoltaics: Theoretical Efficiency Limits. *Appl. Phys. Lett.* **2012**, *101*, 073905.

# Supporting Information

# Understanding Hot-Electron Generation and Plasmon Relaxation in Metal Nanocrystals: Quantum and Classical Mechanisms

Lucas V. Besteiro,[1*] Xiang-Tian Kong,[1,2] Zhiming Wang,[2] Gregory Hartland,[3]

Alexander O. Govorov[1*]

[1] *Department of Physics and Astronomy, Ohio University, Athens OH 45701*

[2] *Institute of Fundamental and Frontier Sciences and State Key Laboratory of Electronic Thin Films and Integrated Devices, University of Electronic Science and Technology of China, Chengdu 610054, China*

[3] *Department of Chemistry and Biochemistry, University of Notre Dame, Notre Dame, IN 46556-5670*

[*] E-mails: lvbesteiro@gmail.com; govorov@ohio.edu

1. **The Drude and interband contributions in the dielectric constant of noble metals**.

The dielectric constant of bulk metals, like Au and Ag, has the following structure [S1]:

$$\varepsilon_{metal}(\omega) = \varepsilon_b + \Delta\varepsilon_{\text{inter-band}}(\omega) + \Delta\varepsilon_{\text{Drude}}(\omega)$$
$$\Delta\varepsilon_{Drude}(\omega) = -\frac{\omega_p^2}{\omega(\omega + i\gamma_{Drude})}, \qquad (S1)$$

where $\Delta\varepsilon_{\text{inter-band}}(\omega)$ is the interband term that is active in gold and silver in the UV and shortwave intervals of the visible light; $\varepsilon_b$ is the screening dielectric constant due to bound atomic-core charges in metals, $\omega_p$ is the bulk plasmon frequency and $\gamma_{Drude}$ is the Drude relaxation constant. The Drude constant is responsible for the friction-like dissipation mechanism in an optically-excited metal NC, where local dissipative electric currents create Joule heat. In our theory, we take the Drude relaxation constant as the rate of momentum relaxation of electron, i.e.

$$\gamma_p = \gamma_{Drude}. \qquad (S2)$$

This assumption is well justified physically, since the long-wavelength limit of the Drude dielectric constant should produce the static conductivity of an electron system; this argument was used and described in ref S2. Moreover, this direct relation (eq S2) follows from the consideration of the decay process of an electric current in an electron plasma [S2]. In the main text we give a table the parameters for the Drude dielectric functions, as well as the Fermi energies and the relaxation times for Au and Ag [S2].

In the models of noble metals, such as Au and Ag, the detailed dielectric function (eq S1) is reduced to the simplified classical Drude function for the long-wavelength intervals, $\lambda > \lambda_{\text{inter-band}}$. In these intervals, the inter-band transitions cannot be excited and the dielectric function has the form

$$\varepsilon_{metal,Drude}(\omega) = \varepsilon_{b,Drude} - \frac{\omega_p^2}{\omega \cdot (\omega + i\gamma_{Drude})}.$$

This function is very convenient for a description of plasmonic effects in the IR region, and in the long-wavelength parts of the visible spectrum. In the main text we also introduced a model of an idealized metal with only intraband transitions, which is regarded as a Drude metal. This model is described by the above function $\varepsilon_{metal,Drude}(\omega)$. These models should be used for the following physical satiations: (1) injection of high-energy electrons to a semiconductor (like Au-$TiO_2$ NCs in [S3]); (2) the experiments with generation of hot electrons in planar metamaterials using red and infrared light [S4,S5]; (3) generation of electrons in hybrid nanostructures (made of Au, Ag and other metals) with plasmonic hot spots [S5].

## 2. Derivation of the energy stored in the plasmonic wave.

According to the textbook [S6], the energy stored in an optically-active, weakly-absorbing media is given by the integral

$$E_{plasmon} = \frac{1}{4\pi} \int dV \frac{d\varepsilon(r,\omega) \cdot \omega}{d\omega} \mathbf{E}_{\omega}(r) \cdot \mathbf{E}_{\omega}^{*}(r), \qquad (S3)$$

where $\mathbf{E}_{\omega}(r)$ is the amplitude of the electric field induced by the surface charges of a NC. We now apply eq S3 to the case of a plasmonic nanosphere. The electric fields inside and outside an optically-driven small sphere are well known [S6]. In particular, here is an expression for the inside electric field:

$$E_{\omega,in} = E_0 \frac{3\varepsilon_0}{2\varepsilon_0 + \varepsilon_{metal}}, \qquad (S4)$$

where $E_0$ is the external, driving field, and $\varepsilon_0$ is the dielectric constant of the matrix. Then, we take the integral in eq S3 and obtain

$$E_{plasmon} = \frac{1}{4\pi}\left(2\varepsilon_0 \left|\frac{\varepsilon_{metal}-\varepsilon_0}{2\varepsilon_0+\varepsilon_{metal}}\right|^2 + (\varepsilon(\omega)+\omega\varepsilon(\omega)')\left|\frac{3\varepsilon_0}{2\varepsilon_0+\varepsilon_{metal}}\right|^2\right)E_0^{2}V_{NP}. \qquad (S5)$$

In the presence an external electric field, the integral (S3) is, of course, infinite; therefore, we involve in the integration only the electric field induced by surface charges and then obtain the results (S5). We should now recall that we are interested in the energy stored in the plasmonic motion in a NC. Since the plasmon has weak dissipation, at this moment we can take the limit $E_0 \to 0$ and neglect the external field, but keep the internal field (S4) nonzero. This is possible because the denominator in eq S4 becomes very small at the plasmon resonance frequency. Then, the plasmon energy (S5) at the plasmonic resonance $\omega = \omega_{p,sphere}$ becomes:

$$E_{plasmon} = \frac{1}{4\pi}\left(2\varepsilon_0 \left|\frac{\varepsilon_{metal}-\varepsilon_0}{3\varepsilon_0}\right|^2 + \left(\varepsilon(\omega)+\omega\varepsilon(\omega)'\right)\right)\left(\mathbf{E}_{\omega,in}\cdot\mathbf{E}_{\omega,in}^*\right)V_{NP}. \quad (S6)$$

In the Drude model, the plasmonic energy stored in a spherical NP is further reduced to

$$E_{plasmon} = \frac{1}{2\pi}V_{NP}\cdot E_{\omega,in}^{\;2}\left(2\varepsilon_0 + \varepsilon_{b,Drude}\right), \quad (S7)$$

where the constants were defined above. To get Eq. S7, we took the limit $\gamma_{Drude} \to 0$ and also assumed $\omega = \omega_{p,sphere}$, where the plasmon frequency follows from the condition $2\varepsilon_0 + \varepsilon_{metal} = 0$.

**3. Derivation of the hot-electron generation rate at a flat surface.** The rate of generation of hot electrons in the flat region of the spectrum (the plateau, $E_F + \delta E_{crit} < \varepsilon < E_F + \hbar\omega$, in Figure S1) can be derived using a model of a slab. We start from the equations for the rates from the main text:

$$Rate_e(\varepsilon) = \frac{dN}{d\varepsilon dt} = \sum_n G_n \cdot P(\varepsilon - \varepsilon_n),$$

$$G_n = \frac{2}{\hbar}\sum_{n'}(f_{n'} - f_n)\left[\left|V_{nn',a}\right|^2 \frac{\gamma_p}{\left(\hbar\omega - \varepsilon_n + \varepsilon_{n'}\right)^2 + \gamma_p^2} + \left|V_{nn',b}\right|^2 \frac{\gamma_p}{\left(\hbar\omega + \varepsilon_n - \varepsilon_{n'}\right)^2 + \gamma_p^2}\right]. \quad (S8)$$

In the slab model [S2,S7]

$$\varepsilon_\mathbf{n} = \frac{\hbar^2 k_\parallel^2}{2m_0} + \frac{\hbar^2 \pi^2 n_z^2}{2m_0 L_z^2},$$

where $L_z$ is the slab width and $k_\parallel^2$ is the in-plane wave vector of an electron. The quantum state of an electron in an extended slab is described by the numbers $n = (k_x, k_y, n_z)$, where one of them is quantized, $n_z = 1, 2, 3, \ldots$. The corresponding electronic wave functions are

$$\psi_n = \sqrt{\frac{2}{L_z}} \frac{1}{\sqrt{A}} \exp[ik_x x] \cdot \exp[ik_y y] \cdot \mathrm{Sin}[\frac{\pi}{L_z} n_z z],$$

where $A$ is the surface area of the slab. Then, the matrix elements in eq S8 can be calculated, considering $P(\varepsilon - \varepsilon_n)$ as a delta function. We also assume that the external filed is normal to the slab. The rate (S8) becomes:

$$Rate_e(\varepsilon) = \frac{2}{\hbar} \sum_{n,n'} (f_{n'} - f_n) \left[ |V_{nn',a}|^2 \frac{\gamma_p}{(\hbar\omega - \varepsilon_n + \varepsilon_{n'})^2 + \gamma_p^2} + |V_{nn',b}|^2 \frac{\gamma_p}{(\hbar\omega + \varepsilon_n - \varepsilon_{n'})^2 + \gamma_p^2} \right] \cdot \delta(\varepsilon - \varepsilon_n).$$

Since the in-plane momentum is conserved,

$$V_{nn',a} = \delta_{k_\parallel, k_\parallel'} \langle \psi_{n_z}(z) | (-ezE_{\omega,z}) | \psi_{n_z'}(z) \rangle.$$

Then, the double sum becomes simpler:

$$Rate_{e,slab}(\varepsilon) = \frac{2}{\hbar} \sum_{n_z, n_z', k_\parallel'} (f_{n'} - f_n) \left[ |V_{n_z n_z',a}|^2 \frac{\gamma_p}{(\hbar\omega - \varepsilon_n + \varepsilon_{n'})^2 + \gamma_p^2} + |V_{n_z n_z',b}|^2 \frac{\gamma_p}{(\hbar\omega + \varepsilon_n - \varepsilon_{n'})^2 + \gamma_p^2} \right] \cdot \delta(\varepsilon - \varepsilon_n)$$

To obtain the high-energy (hot-electron) contribution, we make the following simplifications:

$$\frac{\gamma_p}{(\hbar\omega - \varepsilon_n + \varepsilon_{n'})^2 + \gamma_p^2} \to \pi\delta(\hbar\omega - \varepsilon_n + \varepsilon_{n'}),$$

$$\sum_{n_z, n_z', k_\parallel'} (\ldots) \to 2 \cdot \int_0^\infty dn \int_0^\infty dn' \int_0^\infty \frac{A}{(2\pi)^2} d^2\mathbf{k} \cdot (\ldots).$$

Since the resulting triple integral contains two delta functions, it can be easily taken as a 1D integral. Then, at low temperature ($E_F, \hbar\omega \gg k_B T$) and for a large Fermi energy $E_F > \hbar\omega$, we get

$$Rate_{e,slab}(\varepsilon) = 2 \cdot \frac{2}{\pi^2} \times \frac{e^2 E_F^2}{\hbar} \frac{1}{(\hbar\omega)^4} |E_{\omega,z}|^2 \cdot A. \quad (S9)$$

This is the result for a slab with an area $A$, in the presence of the time-oscillating normal field $E_{\omega,z}$. For our geometry, the rate (S9) describes the generation of hot electrons at two interfaces. For one interface with an area $ds$, we then have [S8]

$$Rate_{e,ds}(\varepsilon) = \frac{2}{\pi^2} \times \frac{e^2 E_F^2}{\hbar} \frac{1}{(\hbar\omega)^4} |E_{\omega,normal}|^2 \cdot ds, \quad E_F + \delta E_{crit} < \varepsilon < E_F + \hbar\omega.$$

In the next step, we integrate this local rate over the whole surface area of a NC with a smooth shape:

$$Rate_{e,NC}(\varepsilon) = 2 \cdot \frac{2}{\pi^2} \times \frac{e^2 E_F^2}{\hbar} \frac{1}{(\hbar\omega)^4} \int_{S_{NC}} |E_{normal}(\theta,\varphi)|^2 \, ds, \quad E_F + \delta E_{crit} < \varepsilon < E_F + \hbar\omega. \quad (S10)$$

For the electron states below the Fermi level, we, of course, obtain an opposite value for the rate:

$$Rate_{e,NC}(\varepsilon) = \frac{dN}{d\varepsilon dt} = -2 \cdot \frac{2}{\pi^2} \times \frac{e^2 E_F^2}{\hbar} \frac{1}{(\hbar\omega)^4} \int_{S_{NC}} |E_{normal}(\theta,\varphi)|^2 \, ds, \quad E_F - \hbar\omega < \varepsilon < E_F - \delta E_{crit}.$$

At this point, the total rate of generation of high energy electrons can be obtained by integrating the value $\frac{dN}{d\varepsilon dt}$ over the energy interval of hot electrons, $E_F + \delta E_{crit} < \varepsilon < E_F + \hbar\omega$. Then, we obtain

$$Rate_{high-energy} = \frac{2}{\pi^2} \times \frac{e^2 E_F^2}{\hbar} \frac{\hbar\omega - \delta E_{crit}}{(\hbar\omega)^4} \int_{S_{NC}} |E_{normal}(\theta,\varphi)|^2 \, ds. \quad (S11)$$

An approximation of this equation was used in the main text as eq 20. Finally, using the rate (S10), we can obtain the energy dissipation for the quantum surface-scattering mechanism integrating the following over the intervals of the hot electrons [S9]:

$$Rate_{high-energy} = \int_{\substack{E_F+\delta E_{crit}<\varepsilon<E_F+\hbar\omega \\ E_F-\hbar\omega<\varepsilon<E_F-\delta E_{crit}}} Rate_{e,NC}(\varepsilon)\cdot\varepsilon\cdot d\varepsilon \approx \int_{\substack{E_F+\delta E_{crit}<\varepsilon<E_F+\hbar\omega \\ E_F-\hbar\omega<\varepsilon<E_F-\delta E_{crit}}} Rate_{e,NC}(\varepsilon)\cdot(\varepsilon-E_F)\cdot d\varepsilon$$

Since the function $Rate_{e,NC}(\varepsilon)$ is approximately flat in the two integration regions, we obtain an equation for the total quantum dissipation in a NC:

$$Q_{hot-electrons} = \frac{2}{\pi^2}\times\frac{e^2 E_F^2}{\hbar}\frac{(\hbar\omega)^2-\delta E^2}{(\hbar\omega)^4}\int_{S_{NC}}|E_{normal}(\theta,\varphi)|^2\times ds \approx$$
$$\approx \frac{2}{\pi^2}\times\frac{e^2 E_F^2}{\hbar}\frac{1}{(\hbar\omega)^2}\int_{S_{NC}}|E_{normal}(\theta,\varphi)|^2\times ds \quad (S12)$$

Then, we can compare eqs (S11) and (S12) to obtain one useful relation given in the main text (see eq 16 and the first equation in Section 5.8):

$$Q_{hot-electrons} = \frac{(\hbar\omega^2-\delta E_{crit}^2)}{(\hbar\omega-\delta E_{crit})}\cdot Rate_{high-energy} \approx \hbar\omega\cdot Rate_{high-energy}$$

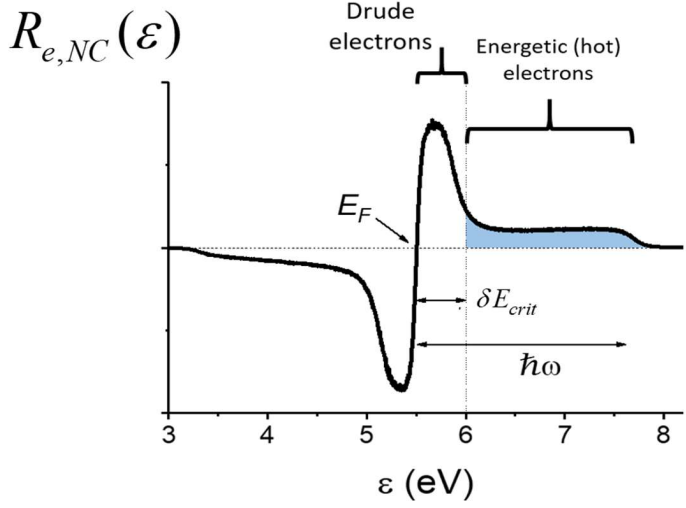

**Figure S1:** Generic distribution of non-equilibrium plasmonic electrons under CW illumination in a NC with smooth surface. The blue area indicates the interval of integration for the hot electrons.

**4. Ratio for the high-energy and low-energy rates.** Eq 55 in the main text can be derived from the following simple arguments. The ratio between the dissipations reads

$$\frac{Q_{high-energy}}{Q_{low-energy,Drude}} = const \cdot \frac{l_{mfp}}{a_0},$$

Then, since the dissipations are given by the energy integrals and we have the estimates

$$Q_{high-energy} \sim Rate_{high-energy} \cdot \hbar\omega,$$

$$Q_{low-energy,Drude} \sim Rate_{low-energy,Drude} \cdot \hbar\Delta k \cdot v_F \sim Rate_{low-energy,Drude} \cdot \hbar\frac{\pi}{a_0} \cdot v_F,$$

$$\Delta k \sim \hbar\frac{\pi}{a_0}.$$

The low-energy interval of transitions in small NCs is governed by dipole-like transitions, which produce the Drude dissipation and, therefore, the change in the electron momentum was taken above as $\Delta k \sim \hbar\frac{\pi}{a_0}$. Then, we combine the above equations and obtain the relation (eq 55)

$$\frac{Rate_{high-energy}}{Rate_{low-energy,Drude}} = const' \cdot \frac{l_{mfp}}{a_0} \frac{v_F/a_0}{\omega}.$$

This derivation explains why the ratios for rates and dissipations are different in magnitude. The reason is the factor $\frac{\hbar v_F / a_0}{\hbar \omega} \ll 1$ for NCs with sizes in the typical experimental range.

## 5. Derivation for the quantum yield for the generation of over-barrier electrons.

The hot-electron parameters of a plasmonic NC (the quantum parameter of plasmon and the quantum yield of over-barrier electrons) are defined as

$$QP_{plasmon} = \frac{Q_{quantum}}{Q_{tot}},$$

$$QY_{plasmon} = \frac{Rate_{hot-electrons,\varepsilon > \Delta E_{barrier}}}{Rate_{tot,\,photon\,absorption}}.$$

The integrated rates of generation for all and over-barrier electrons read:

$$Rate_{electrons} = \int_{E_F}^{\infty} R_e(\varepsilon) d\varepsilon,$$

$$Rate_{hot-electrons,\varepsilon > \Delta E_{barrier}} = \int_{\Delta E_{bar}}^{\infty} R_e(\varepsilon) d\varepsilon.$$

Then, we observed that the spectral rate for high-energy electrons is constant at the plateaus. Therefore, we can easily integrate the above equations and we get

$$Rate_{hot-electrons} \approx \hbar\omega \cdot R_{e,hot-electrons,plateau}$$
$$Rate_{hot-electrons,\varepsilon > \Delta E_{barrier}} \approx (\hbar\omega - \Delta E_{barrier}) \cdot R_{e,hot-electrons,plateau}$$

where $R_{e,hot-electrons,plateau}$ is the spectral rate at the plateaus; for this rate, we have a simple analytical equation in the text for a NC of arbitrary shape, assuming that the surface is smooth. Then, we note that the quantum parameter of plasmon is given by

$$QP_{plasmon} = \frac{\hbar\omega \cdot R_{e,hot-electrons}}{Q_{tot}}.$$

Simultaneously, from the above equations, we obtain:

$$QY_{plasmon} = \frac{\hbar\omega \cdot (\hbar\omega - \Delta E_{barrier}) \cdot R_{e,hot-electrons,plateau}}{Q_{tot}} = \frac{(\hbar\omega - \Delta E_{barrier}) \cdot R_{e,hot-electrons}}{Q_{tot}}$$

Then, we derive the simple connection between $QY_{plasmon}$ and $QP_{plasmon}$:

$$QY_{plasmon} \approx QP_{plasmon} \frac{(\hbar\omega - \Delta E_{barrier})}{\hbar\omega}.$$

**6. The effect of averaging in computing the energy distributions.**

As we mentioned above, the averaging (eq 46) is crucial to obtain reliable and comprehensible results for the energy distributions of plasmonic electrons. The averaging plays two roles: (1) It models the polydispersity of NC sizes in a solution in a typical experiment. (2) It removes strong oscillations in the results as a function of the optical energy and the NC size. After performing the averaging, all features and physical dependences are smooth and transparent for interpretation. Of course, the averaging significantly increases the computational demands when calculating large NCs with hot spots, but it is worth performing.

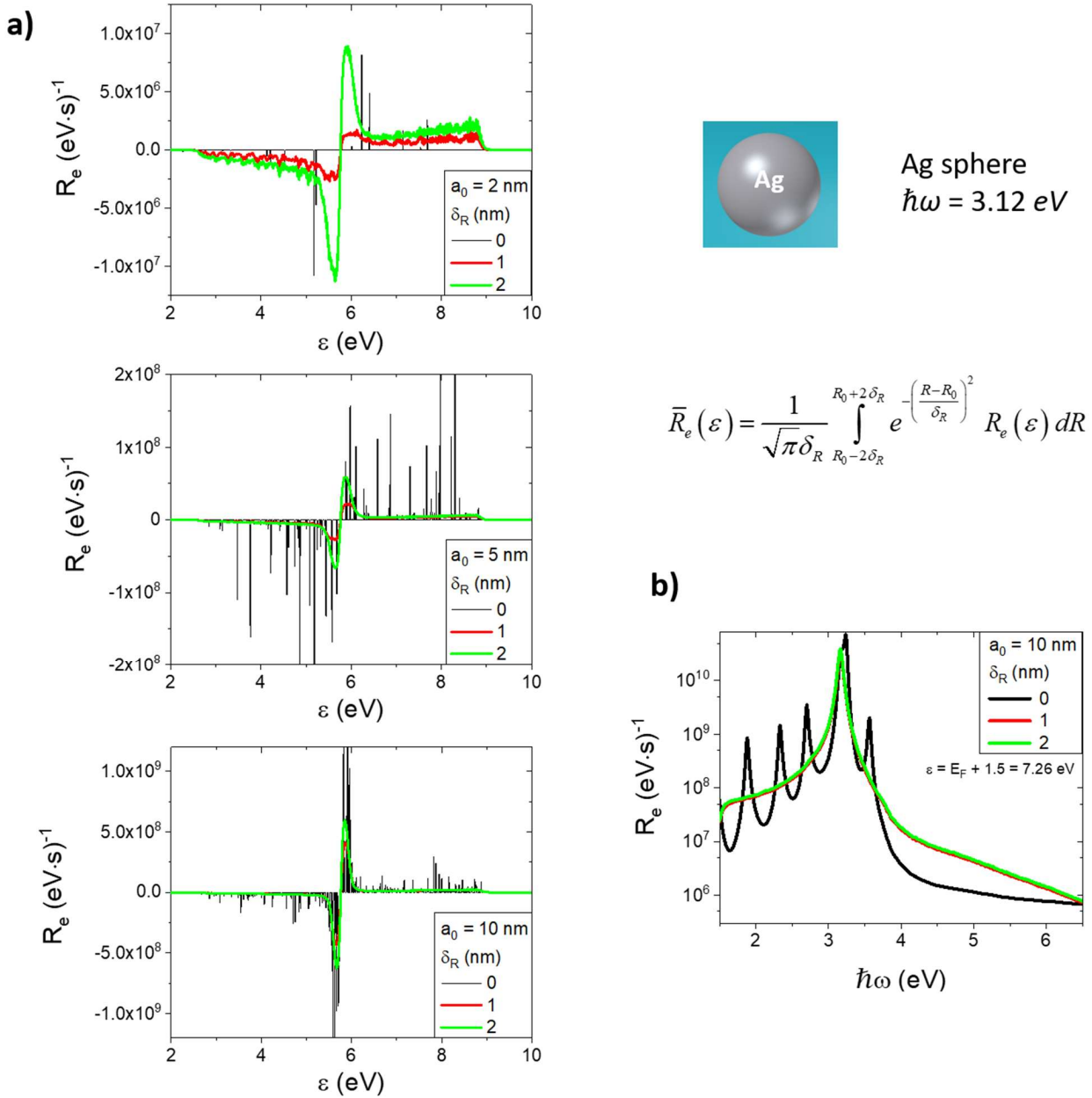

**Figure S2.** Generation rates for three Ag NC sizes, calculated for different averaging parameters. **(a)** Generation rates as a function of the electron energy for the fixed photon energy of 3.12eV. One can see that the averaging is crucial to see the physical properties of the energy distribution. **(b)** The rate as a function of the photon energy for a fixed electron energy taken well above the Fermi level. The plot, given

in semi-logarithmic scale, shows that the averaging is crucial to recognize the physical behaviors of the rates. Insets: Model of the Ag NC and the averaging formula for the rates.

## 7. Definition of the critical energy and generation of the Drude electrons

Figure S3 below shows the definition and the properties of the critical energy, $\delta E_{crit}$, in the spectrum of generation of intraband hot electrons. In the following Figure S4, we show the Drude limit (the limit of large NC sizes) for generation of hot electrons when the majority of photo-excited electrons in a plasmonic wave have small excitation energies. The energy dissipation in this case is given by the classical Drude model.

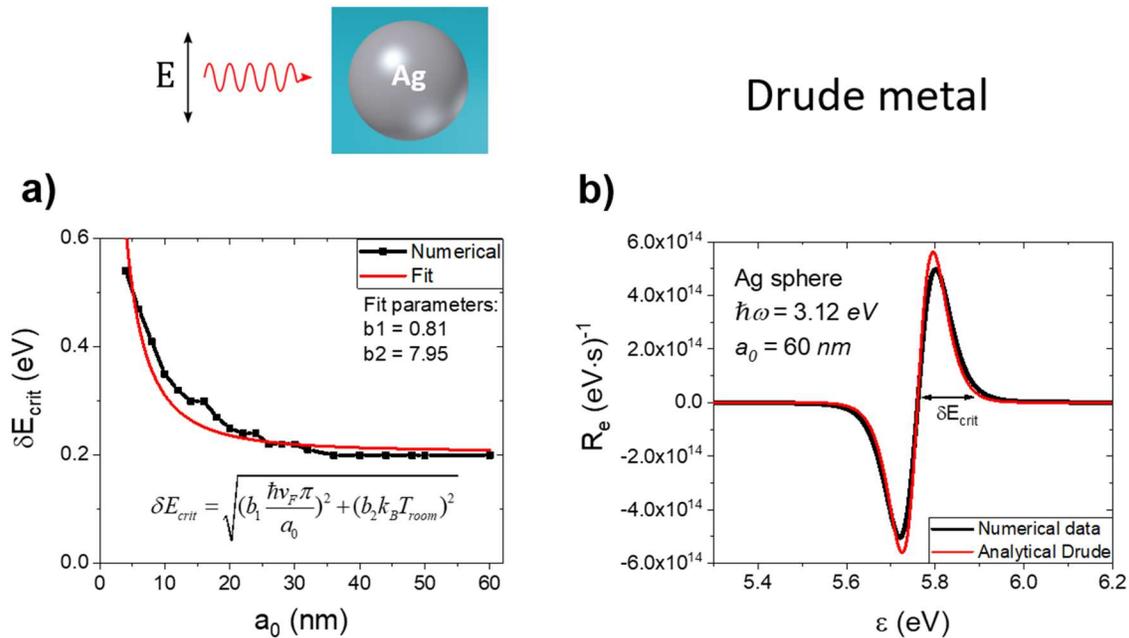

**Figure S3.** (a) Critical energy that sets up the boundary between the Drude and hot electrons. As an example, we show this energy for the case of Ag NCs. (b) Rate of electron generation for a large-size NC and the corresponding analytical results derived from eq 53 in the main text.

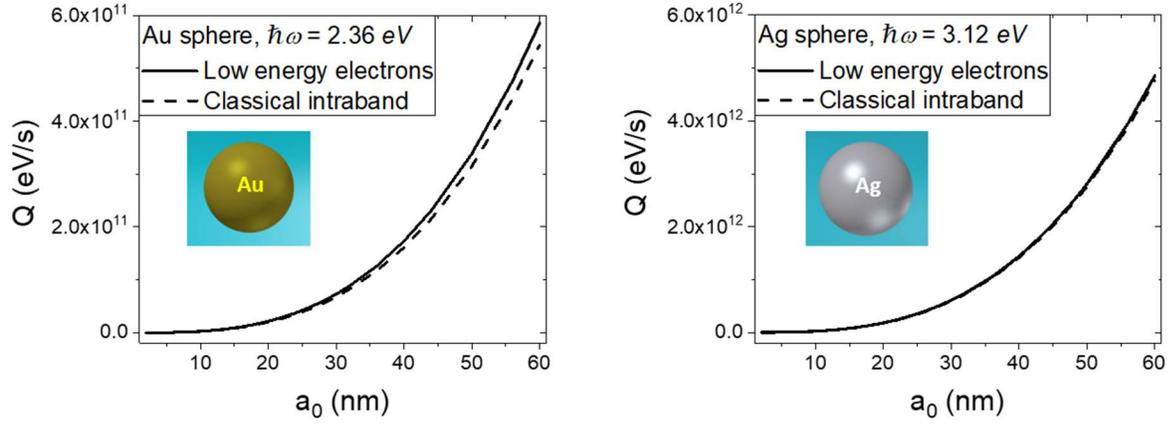

**Figure S4.** Dissipations due to the Drude electrons calculated from the quantum formalism (eq 35) and using the classical Drude equation (eq 53). The incident flux is $I_0 = 3.6 \cdot 10^3 W/cm^2$.

## 8. Rate of plasmon decay due to collisions with the surface in Silver.

The main text shows the data for the plasmonic decay in a spherical Au NC. Here we show similar results for silver as Figure S5.

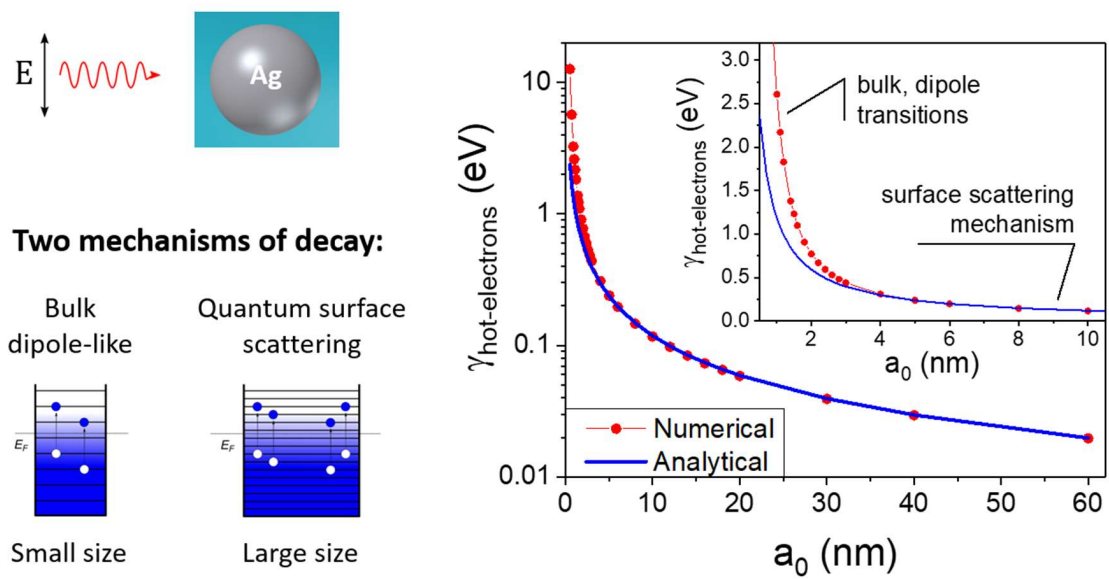

**Figure S5.** Rate of plasmon decay due to collisions with the surface, shown in semi-logarithmic scale and data for small NCs in linear scale in the inset; $\hbar\omega = 3.12\ eV$. Diagrams: Model of an Ag NC and two mechanisms of plasmon decay.

## 9. The interband carriers in Au NCs: The energy intervals for the d-holes and sp-electrons.

The spectra of generation of hot electrons and holes in the sp- and d-bands of gold and silver should be derived, of course, from atomistic calculations. We now use some data from such atomistic, LCAO-type calculations [41]. Figure S6 below shows the intervals where the interband hot electrons and holes become generated.

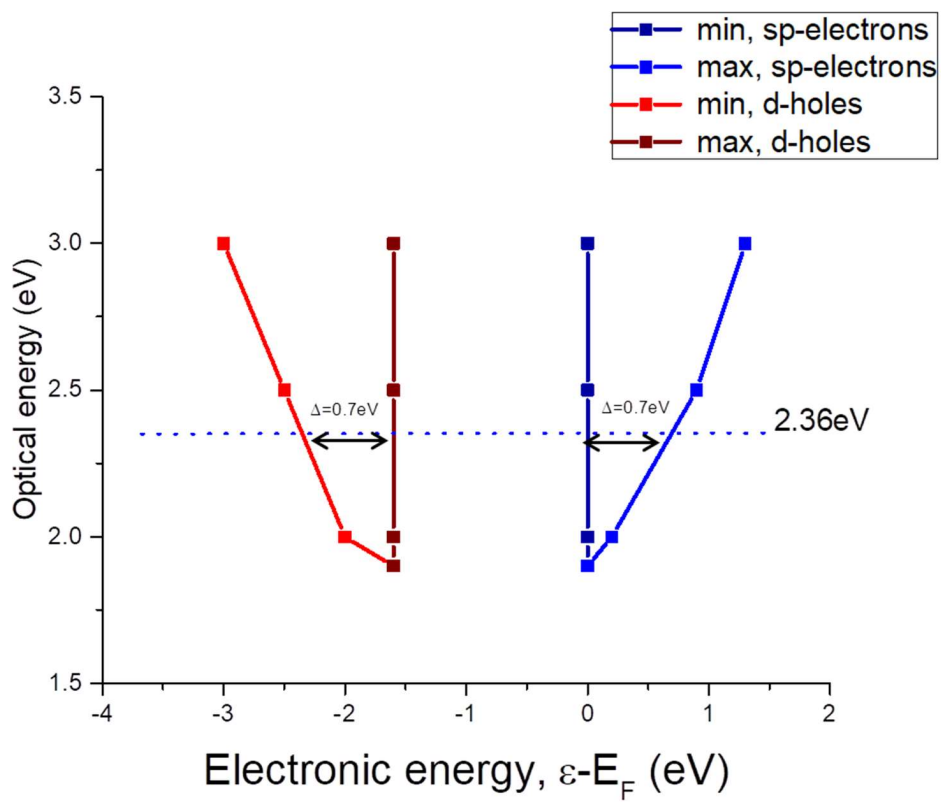

**Figure S6:** Intervals of generation of hot interband carriers in Au NCs are shown on the horizontal axis. The optical energy, $\hbar\omega$, is given on the vertical axis. For a given energy of photons, we see two intervals of hot carriers, just above the Fermi level and at ~ 2.eV below the Fermi level (d-holes). For the plasmon resonance energy of an Au NP (~ 2.36eV), these intervals have extensions of ~0.7eV. These numerical data are taken from Ref 41 and they are also consistent with the interband part of the dielectric function of gold.

**10. Implications of quantum kinetics for time-resolved hot-electron experiments.**

Although the phenomenological picture of plasmon decay shown in Figure 2a is very convenient to analyze and describe photocatalytic, time-resolved and photo-current experiments, there is another kinetic representation that is often used, especially in time-resolved studies (see refs S10 and S11). Figure S7 illustrates it. Electrons are excited from the continuum of occupied states via two types of transitions, low energy (frictional) and high energy (quantum) (Figure S7a). In a fs-pulse experiment, an initial distribution of excited carriers contains some number of high-energy electrons that relax through e-e and e-phonon pathways. Current literature typically depicts short-lived hot-electrons as two flat regions (Figure S7b), while the many-body theory [S2] produces somewhat different transient distributions (Figure S7c). Since the frictional transitions are very active (strong Drude currents), the distribution of excited electrons during a fs-pulse should have a large number of low-energy electrons and some number of hot electrons, as shown in Figure S7c. The details and differences are still under discussion in the literature and should be considered further.

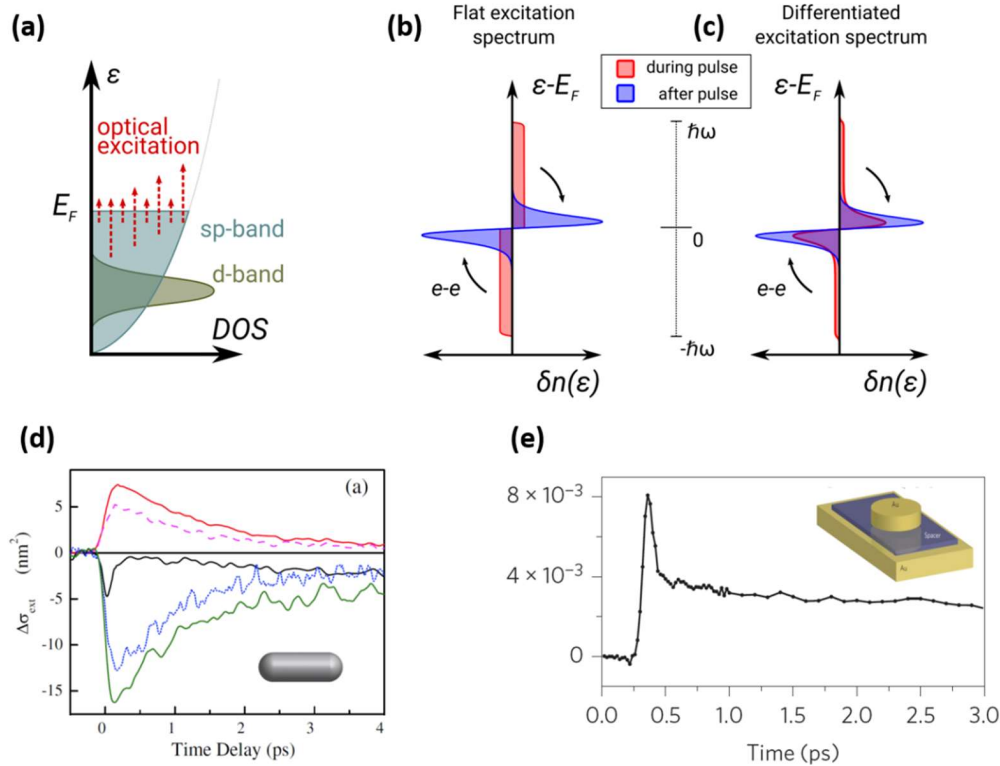

**Figure S7.** Illustrations of the dynamical processes. Panel **(a)** shows optical transitions with low and high excitation energies in the Fermi sea of electrons. **(b,c)** Two scenarios for the temporal electronic populations in a NC excited by a short fs pulse. The qualitative picture shown in (b) is commonly seen in current literature, while the picture in (c) is based on the theory discussed in this paper; our theory involves both quantum kinetics and optical physics. In the picture in panel (c), the distribution of hot electrons has low-energy Drude electrons already during the fs-excitation pulse. **(d,e)** Experimental data for time-resolved plasmonic dynamics of gold nanorods and a planar metastructure with hot spots. The ultra-short, 50 fs, dynamics in Figure (e) indicates the presence of a large number of high-energy (hot) electrons in the metastructure. This is consistent with our results, in which hot spots are able to actively generate hot electrons in large quantities. Panel (d) reproduced with permission from ref S10. Copyrighted by the American Physical Society. Panel (e) reproduced with permission from ref S5. Copyrighted by Nature.